 \documentclass[twocolumn,showpacs]{revtex4}

\usepackage[latin1]{inputenc}
\usepackage{graphicx}%
\usepackage{dcolumn}
\usepackage{amsmath}

\makeatletter
\def\btt#1{\texttt{\@backslashchar#1}}%
\DeclareRobustCommand\bblash{\btt{\@backslashchar}}%
\makeatother

\begin{document}

 \title{Statistical Physics of
Synchronized Traffic Flow: Spatiotemporal Competition between S$\rightarrow$F and S$\rightarrow$J   Instabilities}

 \mark{A Probabilistic Theory of
Synchronized Traffic Flow}

\author{Boris S. Kerner$^1$}

 \affiliation{$^1$
Physics of Transport and Traffic, University Duisburg-Essen,
47048 Duisburg, Germany}


\begin{abstract} 
We have revealed statistical physics of synchronized traffic flow that is governed
by a spatiotemporal competition between    
S$\rightarrow$F and S$\rightarrow$J instabilities (where F, S, and J denote, respectively,
the free flow, synchronized flow, and wide moving jam traffic phases). A probabilistic
analysis of synchronized flow   based on simulations of a
cellular automaton  model in the framework of three-phase
traffic theory is made. This probabilistic analysis
shows that there is a finite range of the initial space-gap between vehicles in synchronized flow
within which  during a chosen time for traffic observation
either synchronized flow persists with probability $P_{\rm S}$,
  or    an S$\rightarrow$F transition  occurs
  with probability $P_{\rm SF}$,
or  else   an S$\rightarrow$J transition occurs
  with probability $P_{\rm SJ}$.
Space-gap dependencies of the probabilities
$P_{\rm S}$, $P_{\rm SF}$, and $P_{\rm SJ}$ have been found.
It has been also found that (i) an initial S$\rightarrow$F   instability
can lead to  sequences of S$\rightarrow$F$\rightarrow$S$\rightarrow$J transitions;
(ii) an initial S$\rightarrow$J   instability
can lead to sequences of  S$\rightarrow$J$\rightarrow$S$\rightarrow$F transitions. 
Each of the phase transitions in the sequences S$\rightarrow$F$\rightarrow$S$\rightarrow$J transitions
and  S$\rightarrow$J$\rightarrow$S$\rightarrow$F transitions   exhibits the nucleation nature;   these sequences
of phase transitions
determine spatiotemporal features of traffic patterns resulting from
the competition between    
S$\rightarrow$F and S$\rightarrow$J instabilities.
The   statistical features of synchronized flow
found for a homogeneous road remain qualitatively for a road with a bottleneck.
However, rather than nuclei for
 S$\rightarrow$F and S$\rightarrow$J instabilities 
occur at random road locations of the homogeneous road, due to
a permanent non-homogeneity introduced by the bottleneck, nuclei for initial
S$\rightarrow$F and S$\rightarrow$J instabilities
appear mostly at the bottleneck.
 \end{abstract}

\maketitle

 \section{Introduction  \label{Int}} 

\begin{figure}
\begin{center}
\includegraphics*[width=8 cm]{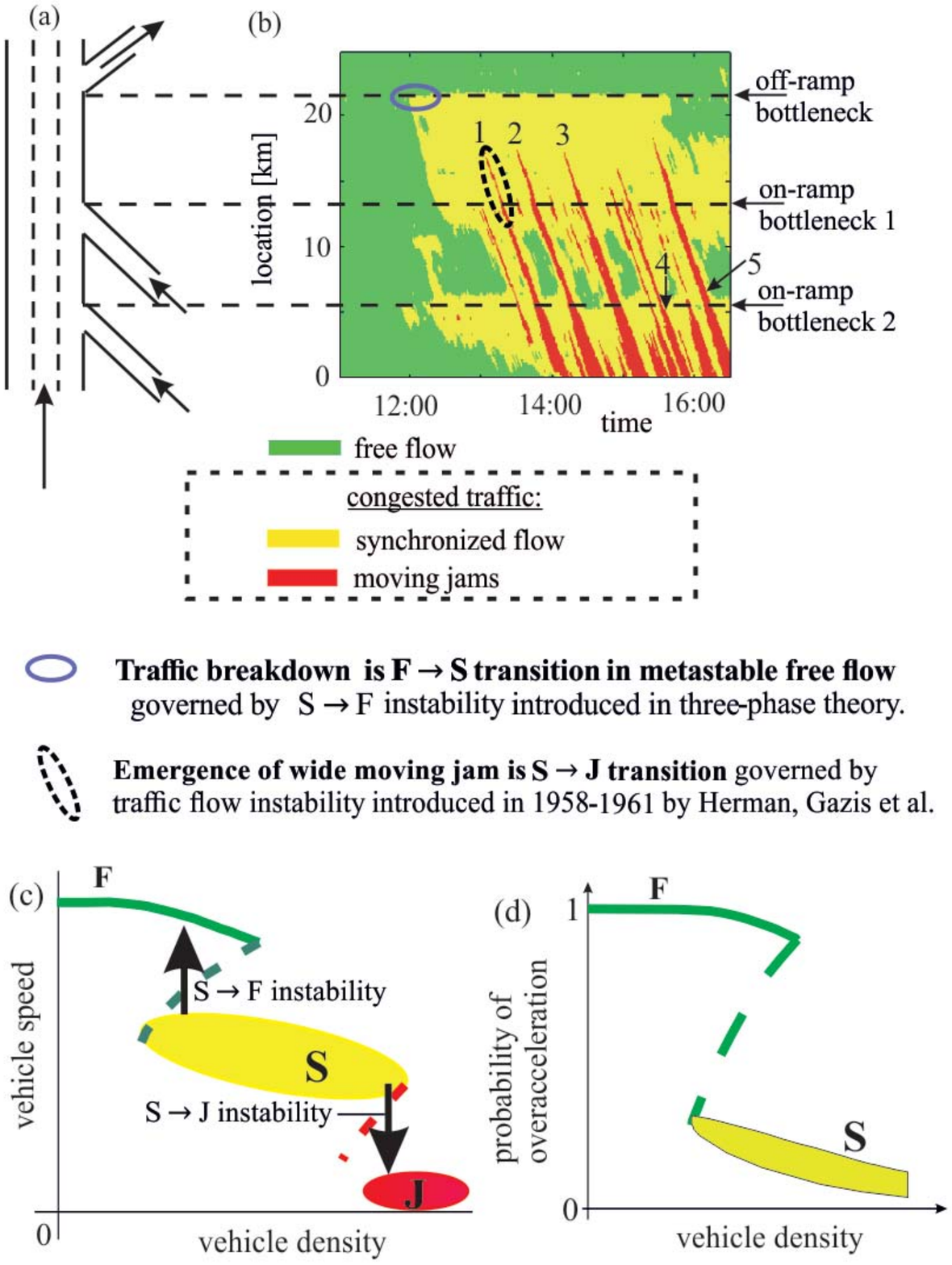}
\end{center}
\caption[]{Explanation of two traffic flow instabilities 
of three-phase   theory~\cite{KernerBook,KernerBook2,KernerBook3}:
(a, b) Typical empirical example of a fragment of complex congested traffic pattern measured
on a highway section with three highway bottlenecks (highway A5-North in Germany)   on March 23, 2001;
(a) sketch of highway section 
(off-ramp bottleneck and two on-ramp bottlenecks); (b)
empirical speed data  
presented in space and time with   averaging 
  method described  in Sec.~C.2 of~\cite{KernerRSch2013}.
 (c) Hypothesis of three-phase theory about    phase transitions in traffic flow: 2Z-characteristic for phase transitions~\cite{KernerBook,Kerner1999B}.
(d) Hypothesis of three-phase theory about discontinuous character of over-acceleration probability~\cite{Kerner1999A,Kerner1999B,Kerner1999C,KernerBook,KernerBook2}.
  F -- free flow phase, S -- synchronized flow phase, J -- wide moving jam phase.  
  }
\label{23032001_North2} 
\end{figure}

Vehicle traffic occurs in space and time. Empirical traffic data measured in space and time
  shows that well-known empirical moving   jams~\cite{Edie,Koshi,Treiterer} 
	occur in free flow through a sequence
of F$\rightarrow$S$\rightarrow$J  phase transitions of three-phase traffic theory (three-phase theory for short)
(F -- free flow, S -- synchronized flow, J -- wide moving jam)~\cite{Kerner1998B,Kerner1999A}: Firstly,  traffic breakdown in free flow at a bottleneck occurs that is a phase transition from free flow  to synchronized flow  (F$\rightarrow$S transition).
Later and usually at a different road location a  phase transition from synchronized flow to a wide moving jam (J)
can be realized (S$\rightarrow$J transition). A typical empirical example of a such sequence of
F$\rightarrow$S$\rightarrow$J   transitions is shown in Fig.~\ref{23032001_North2} (a, b).
Features of empirical F$\rightarrow$S and S$\rightarrow$J   transitions between the   three traffic phases F, S, and J (Fig.~\ref{23032001_North2} (b))
have been explained in the three-phase  theory~\cite{Kerner1998B} by 
the existence two qualitatively different instabilities in the synchronized flow traffic phase: S$\rightarrow$F instability
 and S$\rightarrow$J instability (Fig.~\ref{23032001_North2} (c, d))~\cite{KernerBook,KernerBook2,KernerBook3}.

  The main reason of the three-phase theory is the explanation of the empirical nucleation nature of traffic breakdown (F$\rightarrow$S transition) at the bottleneck. To reach this goal, in congested traffic a new traffic phase called synchronized flow has been introduced~\cite{Kerner1998B,Kerner1999A}. The basic feature of the synchronized flow traffic phase formulated in the three-phase theory leads to the nucleation nature of the F$\rightarrow$S transition. 
In this sense, the synchronized flow traffic phase, which ensures the nucleation nature of the F$\rightarrow$S transition at a highway bottleneck, and the three-phase traffic theory can be considered synonymous.

An S$\rightarrow$J instability is a growing wave of a local  {\it speed   decrease}  
    in synchronized flow. As shown in~\cite{KernerBook}, the
		S$\rightarrow$J instability is associated with  the classical traffic flow instability
	introduced and developed in
  1958--1961 by Herman, Gazis,  Montroll, Potts,   Rothery, and Chandler~\cite{GH195910,Gazis1961A10,GH10,Chandler} from General Motors (GM) Company.  
	The classical traffic flow instability is associated with   a driver over-deceleration effect:
If a vehicle begins
to decelerate unexpectedly, then due to a finite driver reaction time   the following vehicle  
 starts deceleration with a delay.    As a result,
  the   speed of the following vehicle  becomes lower than the speed of the preceding vehicle. If this over-deceleration effect
is realized for   following drivers, the traffic flow instability occurs.  
With the use of very different mathematical approaches, the classical traffic flow instability 
has been incorporated in a huge number of traffic flow models
(see, e.g.,~\cite{Newell1961,Newell1963A,Newell1981,Newell_Stoch,Gipps,Gipps1986,Wiedemann,Whitham1990,ach_Pay197110,ach_Pay197910,Stoc,Bando1995,ach_Kra10,fail_Nagatani1998A,fail_Nagatani1999B,ach_Helbing200010,ach_Aw200010,ach_Jiang2001A} and reviews~\cite{Reviews,Reviews2,Kerner_Review}). All these 
different traffic flow models
 can be considered belonging to the same GM model class. Indeed,
 as found firstly in 1993--1994~\cite{KK1993}, in all these very different traffic flow models 
the classical instability 
 leads to a moving jam (J) formation in free flow (F) (F$\rightarrow$J transition) 
 (see references in~\cite{Reviews2,KernerBook,KernerBook2,KernerBook3,Kerner_Review}). 
 The classical instability  of the GM model class   should explain traffic breakdown, i.e.,
a transition from free flow to congested traffic observed in real 
traffic~\cite{GH195910,Gazis1961A10,GH10,Chandler,Newell1961,Newell1963A,Newell1981,Newell_Stoch,Gipps,Gipps1986,Wiedemann,Whitham1990,ach_Pay197110,ach_Pay197910,Stoc,Bando1995,ach_Kra10,fail_Nagatani1998A,fail_Nagatani1999B,ach_Helbing200010,ach_Aw200010,ach_Jiang2001A,Reviews,Reviews2}).
 However, as shown in~\cite{KernerBook,KernerBook2,KernerBook3,Kerner_Review}, traffic flow models
 models of the GM model class (see references in~\cite{Kerner_Review,KernerBook,KernerBook2,KernerBook3})
failed in the explanation of real traffic breakdown. This is because rather than an F$\rightarrow$J transition
of the models of the GM model class, in all  real field traffic data traffic breakdown is
 an F$\rightarrow$S transition occurring in    metastable free flow
with respect to the F$\rightarrow$S transition~\cite{KR1997,Kerner1997A7,Kerner1998B,Kerner1999A,Kerner1999B,Kerner1999C,KernerBook,KernerBook2,KernerBook3,Kerner_Review,Kerner2002A,Waves,Kerner_Review2}.

 In contrast with the S$\rightarrow$J instability that is a   growing wave of a local  {\it speed   reduction}  
    in synchronized flow, an
	  S$\rightarrow$F instability is a growing wave of a local  {\it speed   increase}  
    in synchronized flow.  
		As shown in~\cite{Kerner2015B}, the S$\rightarrow$F instability exhibits the nucleation nature.
		The nucleation nature of the S$\rightarrow$F instability
		governs the    nucleation nature of the F$\rightarrow$S transition at the bottleneck.
  The S$\rightarrow$F instability     is
associated with  the over-acceleration effect~\cite{Kerner1999A,Kerner1999B,Kerner1999C,KernerBook,KernerBook2,KernerBook3,Kerner_Review}. It is assumed that probability of over-acceleration  
should  
exhibit a   discontinuous character  (Fig.~\ref{23032001_North2} (d)) that is associated with
a driver time-delay in 
over-acceleration~\cite{Kerner1999B,Kerner1999C,KernerBook,KernerBook2,KernerBook3,Kerner_Review}.

 The first mathematical implementation of    hypotheses of the three-phase theory~\cite{Kerner1997A7,Kerner1998B,Kerner1999A,Kerner1999B,Kerner1999C,KernerBook,KernerBook2} has been a stochastic
 continuous in space microscopic model~\cite{KKl} and a  cellular automaton (CA)
  three-phase model~\cite{KKW}, which 
 have been further developed for 
 different applications in~\cite{KKl2003A,Kerner2018A,Kerner2018B,Kerner2008C,Kerner2008D,KKl2009A,Kerner_2014,KKl2004A,Kerner_Hyp,KKHS2013,KKS2014A,Heavy,KKS2011,KKl2006AA,Kerner_EPL,Kerner_Diff}.
 Over time there has been developed 
a number of other 
   traffic  flow
models   (e.g.,~\cite{Davis,Lee_Sch2004A,Jiang2004A,Gao2007,Davis2006,Davis2006b,Davis2006d,Davis2006e,Davis2010,Davis2011,Jiang2007A,Jiang2005A,Jiang2005B,Jiang2007C,Pott2007A,Li,Wu2008,Laval2007A8,Hoogendoorn20088,Wu2009,Jia2009,Tian2009,He2009,Jin2010,Klenov,Klenov2,Kokubo,LeeKim2011,Jin2011,Neto2011,Zhang2011,Wei-Hsun2011IEEEA,Lee2011A,Tian2012,Kimathi2012B,Wang2012A,Tian2012B,Qiu2013,YangLu2013A,KnorrSch2013A,XiangZhengTao2013A,Mendez2013A,Rui2014A,Hausken2015A,Tian2015A,Rui2015C,Rui2015D,Xu2015A,Davis2015A,Qian2017,Tian_En2018,HaifeiYang2018}) that incorporate some of the  hypotheses of the three-phase 
theory~\cite{Kerner1999A,Kerner1999B,Kerner1999C,KernerBook,KernerBook2}.

Separately from each other the S$\rightarrow$F   and S$\rightarrow$J instabilities
have already been studied (for a review see, e.g.,~\cite{KernerBook3,Kerner2018B}).
However, from the hypotheses of the three-phase theory it could be expected that there should
be a competition between
the S$\rightarrow$J and S$\rightarrow$F instabilities in synchronized  traffic flow.  Such a competition
between the two qualitatively different traffic flow instabilities in synchronized flow
 has {\it not} been found up to now.

In this paper, we have revealed that there is indeed a spatiotemporal competition   between
the S$\rightarrow$J and S$\rightarrow$F instabilities.   It has been found that this competition
 effects considerably on  statistical features of vehicular traffic flow.
   Either the S$\rightarrow$J instability~\cite{KernerBook} or the S$\rightarrow$F instability exhibits the nucleation nature~\cite{Kerner2015B}. For this reason,
	to study probabilistic features of the {\it  spontaneous} occurrence of  
	these instabilities as well as their competition, we use a stochastic
traffic flow  model with a relatively large amplitude of model fluctuations. This model feature exhibits  
the KKSW (Kerner-Klenov-Schreckenberg-Wolf) CA (cellular automaton) 
three-phase traffic flow model~\cite{KKW,KKHS2013,KKS2014A}.   Because the  KKSW CA   model 
 has already been published, we present it in Appendix~\ref{App_KKSW}.
Main contributions of this paper are as follows:

(i)  We show that there is a finite range of the initial space-gap between vehicles in synchronized flow
within which  during a chosen time $T_{\rm ob}$ for traffic observation
either synchronized flow persists with probability $P_{\rm S}$,
  or   firstly an S$\rightarrow$F transition  occurs
in synchronized flow with probability $P_{\rm SF}$,
or  else firstly an S$\rightarrow$J transition occurs
in synchronized flow with probability $P_{\rm SJ}$.

(ii)
It has been also found that an   S$\rightarrow$F  transition
can lead to  sequences of S$\rightarrow$F$\rightarrow$S$\rightarrow$J transitions.

(iii) We show that an   S$\rightarrow$J   transition
can lead to sequences of  S$\rightarrow$J$\rightarrow$S$\rightarrow$F transitions.
 
(iv) The sequences of phase transitions of items (ii) and (iii)
determine spatiotemporal features of traffic patterns resulting from
the competition between    
S$\rightarrow$F and S$\rightarrow$J instabilities.

(vi) 
The above statistical features of vehicular traffic
found for a homogeneous road remain qualitatively for a road with an on-ramp bottleneck.
In particular, 
  flow-rate dependencies of the probabilities
$P_{\rm S}$, $P_{\rm SF}$, and $P_{\rm SJ}$ 
for synchronized flow at the bottleneck are qualitatively the same
as  the space-gap dependencies of these probabilities found in the paper for the homogeneous road.
The main difference between the latter two cases is that due to
a permanent non-homogeneity introduced by a bottleneck, nuclei for initial
S$\rightarrow$F and S$\rightarrow$J instabilities
appear mostly at the bottleneck.

  The article is organized as follows. In Sec.~\ref{Probabilistic_S},
	we develop the statistical physics  of the   S$\rightarrow$F  
		and S$\rightarrow$J transitions occurring due to the spatiotemporal competition of
		S$\rightarrow$F and S$\rightarrow$J instabilities in the same initial state of synchronized flow
		on a homogeneous road.
		In particular,  we study
			probabilistic characteristics of the phase transitions and  
			microscopic effects governing the phase transitions.
			In  Sec.~\ref{Probabilistic_On_S},
			these results of the statistical physics of synchronized flow
			are applied for a study of
		traffic phenomena occurring in synchronized flow
		at an on-ramp bottleneck.
		In Discussion, we consider 
the 
effect of the average space gap  between vehicles on congested patterns
	(Sec.~\ref{Space_gap_S}), the evolution of  congested patterns at a bottleneck 
	due to the change in the on-ramp inflow rate
	(Sec.~\ref{Patterns_S}), some peculiarities of
	 synchronized flow that occurs spontaneously
	at the bottleneck after sequences of F$\rightarrow$S$\rightarrow$F transitions
	(Sec.~\ref{FSF_S}) as well as formulate conclusions of the paper.

  \section{Probabilistic features of competition between
	S$\rightarrow$F and S$\rightarrow$J instabilities on circular homogeneous road  \label{Probabilistic_S}}

	 To reveal   
	 statistical features of synchronized flow governed by the competition between
	S$\rightarrow$F and S$\rightarrow$J instabilities,   in 
	Sec.~\ref{Probabilistic_S} we have simulated  a spatiotemporal evolution of  
	synchronized flow on a   circular homogeneous single-lane road
	(road length   is 25 km) (Fig.~\ref{Realizations_Fig}).  
We have made a large number $N_{\rm r}$
	(where $N_{\rm r}\gg 1$) of different simulation realizations (runs) in which    
		a diverse variety of critical traffic spatiotemporal phenomena occur  
  with different probabilities.	   
		Some characteristic realizations
		are presented in Figs.~\ref{Realizations_Fig} (a--g). In each of the realizations
	the initial state of synchronized flow on the whole road is a spatially homogeneous one 
	and it is  the same~\cite{Rand}:
	At time instant $t=0$, all vehicles are located at the same chosen initial space gap
   between vehicles $g_{\rm ini}$;   all vehicles begin
	to move simultaneously at the same initial chosen synchronized flow speed 
	$v^{\rm (syn)}_{\rm ini}$.

\subsection{A diverse variety of critical phenomena found  in different
 realizations simulated at the same model parameters
   \label{Real_Sub}}

	  \begin{figure}
\begin{center} 
\includegraphics*[width=8 cm]{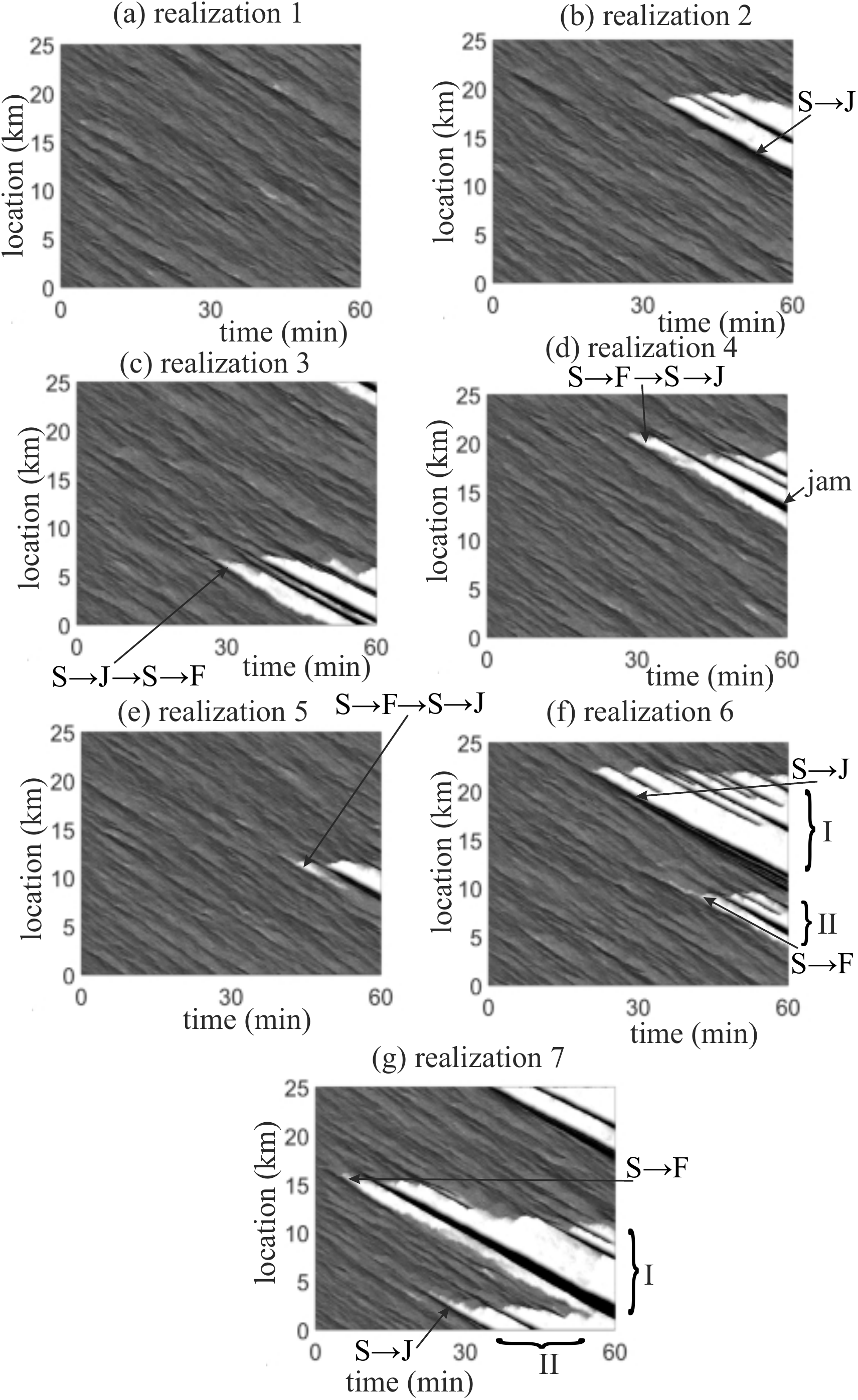}
\end{center}
\caption[]{Characteristic simulation realizations calculated at   the same  
	initial state of synchronized flow  with $g_{\rm ini}=$ 19.5 m,
	$v^{\rm (syn)}_{\rm ini}=$ 15 m/s (54 km/h) and at the same 
	set of model parameters
  of the KKSW CA model~\cite{Rand}.   
 Vehicle  speed data presented by  regions with 
variable shades of gray    (shades of gray vary from white to black when the speed 
decreases from 120 km/h (white) to 0 km/h (black)).
Arrows S$\rightarrow$F label  S$\rightarrow$F transition, arrows S$\rightarrow$J label S$\rightarrow$J transition.
  }
\label{Realizations_Fig} 
\end{figure}
	
	It turns out that the existence of the
diverse variety of  critical traffic   phenomena (Fig.~\ref{Realizations_Fig})
	is indeed associated with a spatiotemporal competition between
	S$\rightarrow$F and S$\rightarrow$J instabilities in synchronized   flow.
 We have found three {\it basic} cases of this  spatiotemporal competition:
\begin{description}
\item (i) Neither the S$\rightarrow$F instability nor the S$\rightarrow$J instability that emerge randomly
on different road locations   leads to S$\rightarrow$F or S$\rightarrow$J transitions.
This is because the development
of each of the instabilities is interrupted over time
(see Sec.~\ref{Waves_Sub} below). As a result, synchronized flow   persists during the whole time
	interval $T_{\rm ob}$ (realization 1 in Fig.~\ref{Realizations_Fig} (a)).
\item (ii) 
 At some random road location, firstly due to the development of
an S$\rightarrow$J instability,  
an S$\rightarrow$J transition occurs (realizations 2, 3, and 6 in Fig.~\ref{Realizations_Fig} (b, c, f)).
\item (iii)
At some random road location, firstly due to the development of
an S$\rightarrow$F instability,   
an S$\rightarrow$F transition occurs (realizations  4, 5, and 7 in Fig.~\ref{Realizations_Fig} (d, e, g)).\end{description}

	\subsection{Probabilities of S$\rightarrow$F and S$\rightarrow$J transitions
as functions  of average space gap between vehicles in synchronized flow \label{Prob_Sub}}

There is a range of 	  the average space gap $\overline g$
(averaged over the circular road)  between vehicles  in the synchronized flow
within which with different probabilities one of the three mentioned basic
 cases (Sec.~\ref{Real_Sub}) of the  spatiotemporal competition
between
	S$\rightarrow$F and S$\rightarrow$J instabilities in synchronized   flow  
	occurs randomly (Fig.~\ref{Prob_Fig}). The average space gap   $\overline g$ is equal  
	to the chosen initial space gap $g_{\rm ini}$.
	Obviously that   the space-gap dependencies of the probability
	$P_{\rm S}(\overline g)$ that synchronized flow persists
	(Fig.~\ref{Realizations_Fig} (a)),  the probability $P_{\rm SJ}(\overline g)$
	that firstly an S$\rightarrow$J transition occurs  
	(Fig.~\ref{Realizations_Fig} (b, c, f)), and the probability $P_{\rm SF}(\overline g)$
	that firstly an S$\rightarrow$F transition occurs
 	(Fig.~\ref{Realizations_Fig} (d, e, g)) during the   time interval $T_{\rm ob}$ satisfy condition: 
		\begin{equation}
P_{\rm S}(\overline g)+P_{\rm SF}(\overline g)+P_{\rm SJ}(\overline g)=1.
\label{Prob2_F_gen}
\end{equation}

  \begin{figure}
\begin{center}
\includegraphics*[width=7.5 cm]{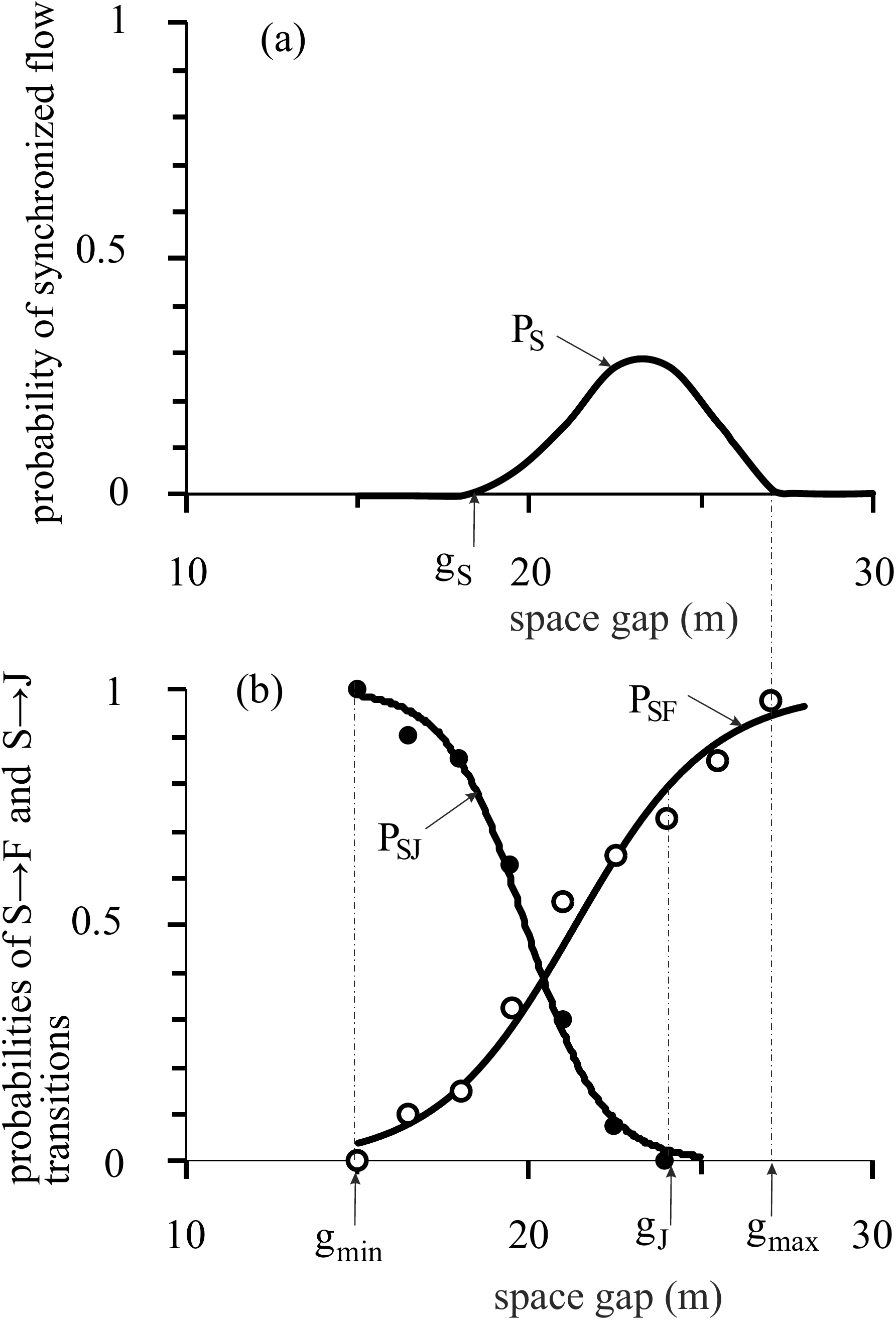}
\end{center}
\caption[]{Probability     $P_{\rm S}(\overline g)$ (curve $P_{\rm S}$)   (a) and
probabilities   $P_{\rm SF}(\overline g)$ (curve $P_{\rm SF}$), 
$P_{\rm SJ}(\overline g)$ (curve $P_{\rm SJ}$) (b)
as functions  of  
$\overline g=g_{\rm ini}$.
For calculation of the space-gap functions of the  
probabilities $P_{\rm SF}(\overline g)$ and $P_{\rm SJ}(\overline g)$,  
at each given value of $g_{\rm ini}$
	different simulation realizations (runs) $N_{\rm r}=$ 40 during the same time interval  
	$T_{\rm ob}=$ 60 min have been made~\cite{Rand}.
 Then, $P_{\rm SF}(\overline g)=n^{\rm (SF)}_{\rm r}/N_{\rm r}$,
	$P_{\rm SJ}(\overline g)=n^{\rm (SJ)}_{\rm r}/N_{\rm r}$, where $n^{\rm (SF)}_{\rm r}$ is the number of realizations in which S$\rightarrow$F transition  has firstly occurred during the time interval $T_{\rm ob}$,
	$n^{\rm (SJ)}_{\rm r}$ is the number of realizations in which S$\rightarrow$J transition  has firstly occurred during the time interval $T_{\rm ob}$.
	Respectively, $P_{\rm S}(\overline g)=1-(P_{\rm SF}(\overline g)+P_{\rm SJ}(\overline g))$.
	Other model parameters are the same as those in Fig.~\ref{Realizations_Fig}.
	Calculated parameters: $g_{\rm min}=15$ m, $g_{\rm S}=18$ m,
	$g_{\rm J}=24$ m,  $g_{\rm max}=27$ m;
	for function $P_{\rm SF}(\overline g)$ (\ref{Prob_SF_circle}) parameters $\alpha=$ 0.52  $\rm m^{-1}$
		and $g_{\rm p}=$ 21.1 m; for function $P_{\rm SJ}(\overline g)$ (\ref{Prob_SJ_circle}) 
		parameters $\alpha=$ 0.9  $\rm m^{-1}$
		and $g_{\rm p}=$ 19.6 m.
  }
\label{Prob_Fig} 
\end{figure}
	
	From a study of $P_{\rm S}(\overline g)$  (Fig.~\ref{Prob_Fig} (a)) we have found that 
	\begin{equation}
P_{\rm S}(\overline g)>0 \quad {\rm only \ if} \    g_{\rm S} < \overline g < g_{\rm max}.
\label{Prob2_F}
\end{equation}
Otherwise, outside the space gap range in (\ref{Prob2_F}) 
\begin{equation}
P_{\rm S}(\overline g)=0 \quad {\rm  if} \  \overline g \leq   g_{\rm S} \ {\rm or} \
 \overline g \geq   g_{\rm max}.
\label{Prob3_F}
\end{equation}
Under condition (\ref{Prob3_F}),
during the observation time $T_{\rm ob}$ 
  either an S$\rightarrow$F transition or an S$\rightarrow$J transition occurs in synchronized flow.

We have found  that the probability of the S$\rightarrow$F transition
$P_{\rm SF}(\overline g)$ is an increasing function of $\overline g$
(curve $P_{\rm SF}$ in Fig.~\ref{Prob_Fig} (b)). Contrarily, 
the probability of the S$\rightarrow$J transition
$P_{\rm SJ}(\overline g)$ is a decreasing function of $\overline g$
(curve $P_{\rm SJ}$ in Fig.~\ref{Prob_Fig} (b))~\cite{Prob_R}.   
There is a range of the average space gap $\overline g$
within which both $P_{\rm SF}(\overline g)>0$ and $P_{\rm SJ}(\overline g)>0$:
\begin{equation}
P_{\rm SF}(\overline g)>0 \ {\rm and} \  P_{\rm SJ}(\overline g)>0 \ {\rm at} \   
 g_{\rm min} < \overline g    <  g_{\rm J},
\label{PP_F}
\end{equation}
where $g_{\rm min}$ and $g_{\rm J}$ some characteristic values of the average space gap
$\overline g$ in synchronized flow (Fig.~\ref{Prob_Fig} (b)). Within the space gap range (\ref{PP_F}), 
randomly either the S$\rightarrow$F transition (Fig.~\ref{Realizations_Fig} (d))
or the S$\rightarrow$J transition (Fig.~\ref{Realizations_Fig} (b)) is possible in  synchronized flow.
Within the space gap range $g_{\rm J}  \leq \overline g < g_{\rm max}$,  $P_{\rm SJ}(\overline g)=0$, therefore,
either an S$\rightarrow$F transition occurs or synchronized flow   persists    (Fig.~\ref{Prob_Fig}).
	
 The   probabilities
$P_{\rm SF}(\overline g)$ (empty circles in Fig.~\ref{Prob_Fig} (b))
and $P_{\rm SJ}(\overline g)$ (black circles in Fig.~\ref{Prob_Fig} (b))
      are well  fitted, respectively, by the functions:
\begin{equation}
P_{\rm SF}(\overline g)=\frac{1}{1+ {\rm exp}[-\alpha(\overline g-  g_{\rm p})]},
\label{Prob_SF_circle}
\end{equation}
\begin{equation}
P_{\rm SJ}(\overline g)=\frac{1}{1+ {\rm exp}[\alpha(\overline g-  g_{\rm p})]},
\label{Prob_SJ_circle}
\end{equation}
where   $\alpha$ and $g_{\rm p}$ are parameters (Fig.~\ref{Prob_Fig}).

		\subsection{Nucleation-interruption effects resulting in
		dissolving speed waves  in synchronized flow:
		S$\rightarrow$F$\rightarrow$S    
		and S$\rightarrow$J$\rightarrow$S  transitions \label{Waves_Sub}}

		In synchronized flow   
		(Fig.~\ref{Realizations_Fig} (a)), due to  random deceleration and acceleration of vehicles, there are many different local speed  disturbances
		(Fig.~\ref{S_circle} (a)). Within   the disturbances the speed is either smaller (local speed reduction) or larger (local speed increase)
		than the average synchronized flow speed denoted by $v^{\rm (syn)}_{\rm av}$.

  \begin{figure}
\begin{center}
\includegraphics*[width=8 cm]{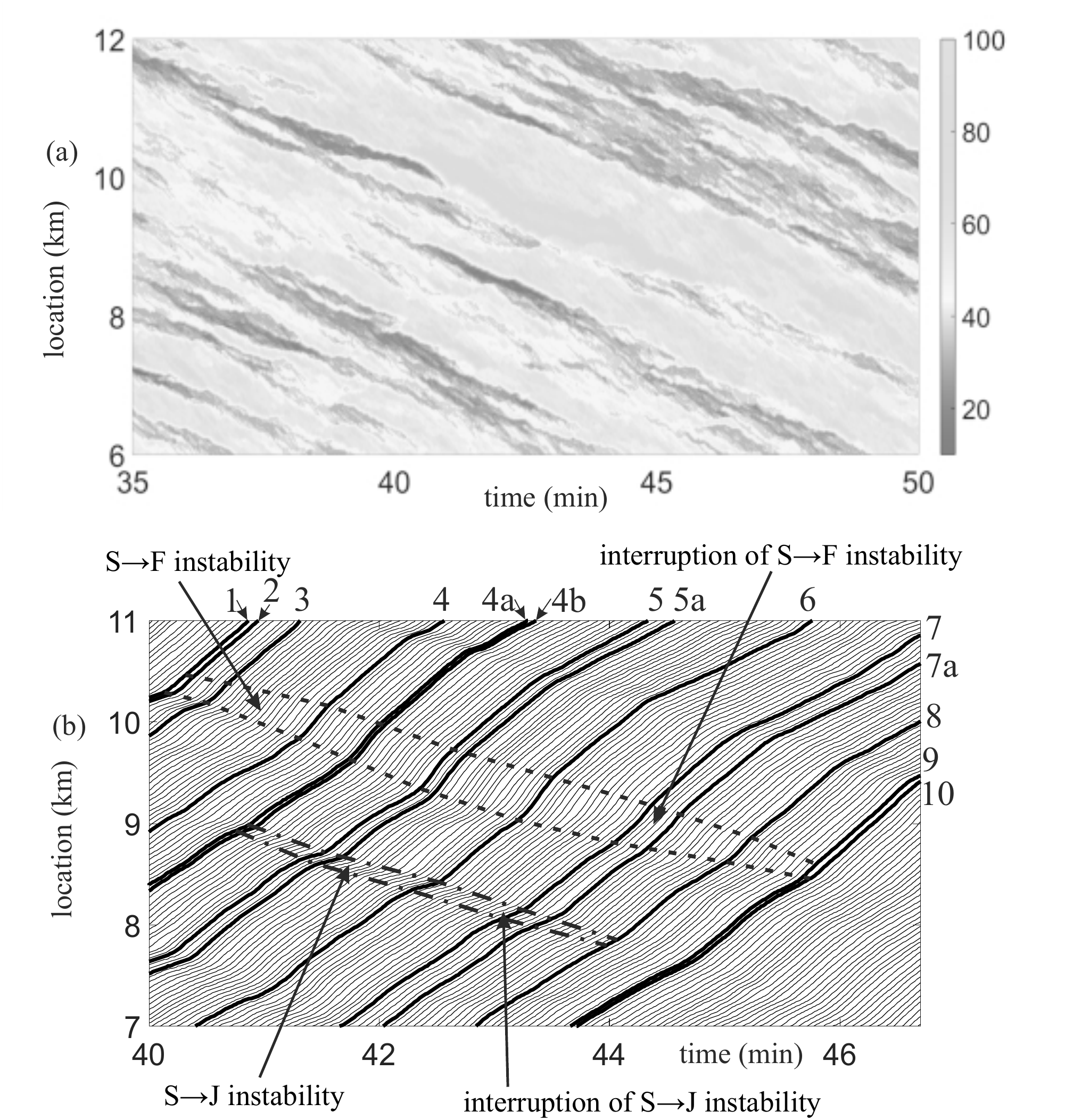}
\end{center}
\caption[]{Continuation of Fig.~\ref{Realizations_Fig} (a).
 Dissolving speed waves  in synchronized flow:
(a) Microscopic speed data presented in space and time.
 (b)  Vehicle trajectories (each 2nd vehicle trajectory is shown).
In (b), a region bounded by dashed curves marks S$\rightarrow$F instability and its interruption;
 a region bounded by dashed-dotted curves marks S$\rightarrow$J instability and its interruption.
  }
\label{S_circle} 
\end{figure}

  \begin{figure}
\begin{center}
\includegraphics*[width=8 cm]{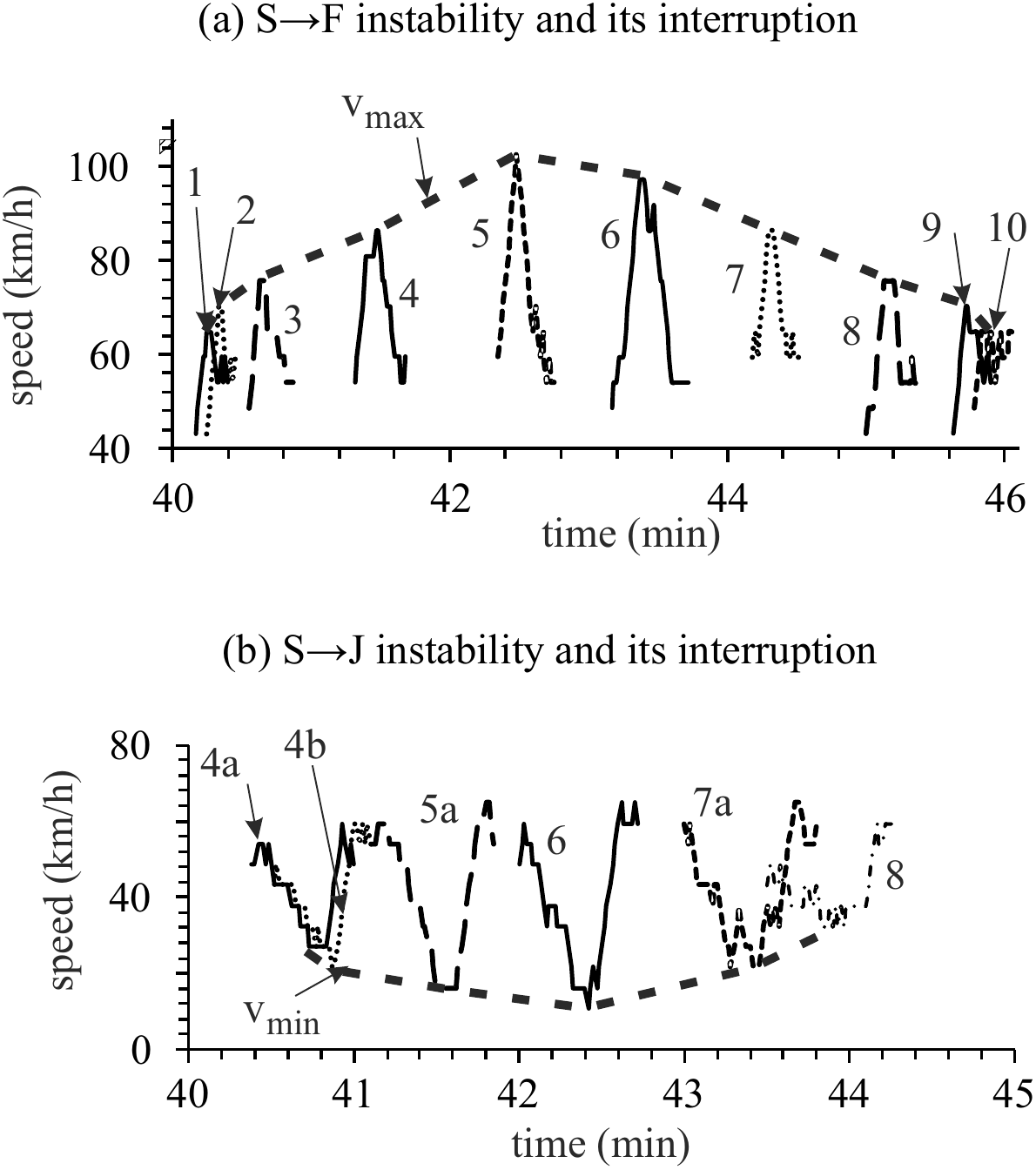}
\end{center}
\caption[]{Continuation of Fig.~\ref{S_circle} (b):
(a) Fragments of microscopic speeds illustrating
S$\rightarrow$F instability and its interruption; bold dashed curve $v_{\rm max}(t)$
shows the time-dependence of the maximum speed on vehicle trajectories. (b) Microscopic speeds illustrating
S$\rightarrow$J instability and its interruption; bold dashed-dotted curve $v_{\rm min}(t)$
shows the time-dependence of the minimum speed on vehicle trajectories.
  Vehicle numbers in (a, b) are the same as those in Fig.~\ref{S_circle} (b).
  }
\label{S_J_F_circle} 
\end{figure}

		If a local speed increase is large enough,  an S$\rightarrow$F instability is realized
		(labeled by $\lq\lq$S$\rightarrow$F instability" in Fig.~\ref{S_circle} (b)):
		The maximum speed inside  a speed wave caused by this local speed increase  begins to grow over time
		 (vehicle trajectories 1--5 in Figs.~\ref{S_circle} (b) and~\ref{S_J_F_circle} (a)).
		
		Respectively, if a local speed decrease is large enough,  an S$\rightarrow$J instability is realized
		(labeled by $\lq\lq$S$\rightarrow$J instability" in Fig.~\ref{S_circle} (b)):
		The minimum speed inside  a speed wave caused by this local speed decrease  begins to decrease over time
		 (vehicle trajectories 4a, 4b, 5a, 6 in Figs.~\ref{S_circle} (b) and~\ref{S_J_F_circle} (b)); the growing speed wave of the local speed decrease in synchronized flow
		resulting from the S$\rightarrow$J instability is also called a growing narrow moving jam~\cite{KernerBook}.

However, in simulation realization 1 under consideration (Fig.~\ref{Realizations_Fig} (a))
due to speed adaptation effect
	that is the same as already explained in~\cite{Kerner2015B}
no phase transitions occur in synchronized flow. This means that the development of
any of the S$\rightarrow$F and S$\rightarrow$J instabilities that  can occur randomly at different road locations are interrupted over time. The interruption of the development of
the S$\rightarrow$F instability   shown in Fig.~\ref{S_J_F_circle} (a)  
(trajectories 6--10) leads to a dissolving speed wave of the local speed increase.
 The   interruption of the development of
the S$\rightarrow$J instability   shown in Fig.~\ref{S_J_F_circle} (b) 
(trajectories 7a and 8) leads to a dissolving speed wave of the local speed decrease, i.e., the dissolution of the narrow moving jam~\cite{KernerBook}. 

In comparison with a known nucleation-interruption 
effect of narrow moving jam emergence and dissolution in synchronized flow~\cite{KernerBook},
 in the case under consideration
due to the competition of the S$\rightarrow$F and S$\rightarrow$J instabilities,  
  two different     nucleation-interruption effects
	interacting with each other in space and time   occur:
	(i) The emergence with the subsequent dissolution of
	the speed wave of the local speed increase (Fig.~\ref{S_J_F_circle} (a)) and (ii) the
	emergence with the subsequent dissolution of
	a narrow moving jam (Fig.~\ref{S_J_F_circle} (b)).

\subsection{S$\rightarrow$J instability initiating S$\rightarrow$F instability:
S$\rightarrow$J$\rightarrow$S$\rightarrow$F transitions \label{S-J-S-F_trans}}

In simulation realization 2 (Fig.~\ref{Realizations_Fig} (b)),  
the development of an S$\rightarrow$J instability is not interrupted over time; therefore,   the development of the S$\rightarrow$J instability leads to an S$\rightarrow$J transition, i.e., wide moving jam emergence. The microscopic development of this effect
can be seen on  trajectories 1--5 in Fig.~\ref{S_J_circle} (a)
and on trajectories 1--3 in Fig.~\ref{S_J_circle_speed} (a)). The S$\rightarrow$J transition 
(Fig.~\ref{S_J_circle_speed} (a)) is a well-known effect~\cite{KernerBook}.
However, due to the possibility of  random occurrence of either S$\rightarrow$F instability
or S$\rightarrow$J instability at the same model parameters, we have revealed
nucleation phenomena that have been unknown up to now (Fig.~\ref{S_J_circle} (b, c)). 

\subsubsection{S$\rightarrow$F instability and its interruption   downstream   of moving jam 
\label{S-J-S-F_trans1}}

  \begin{figure}
\begin{center}
\includegraphics*[width=8 cm]{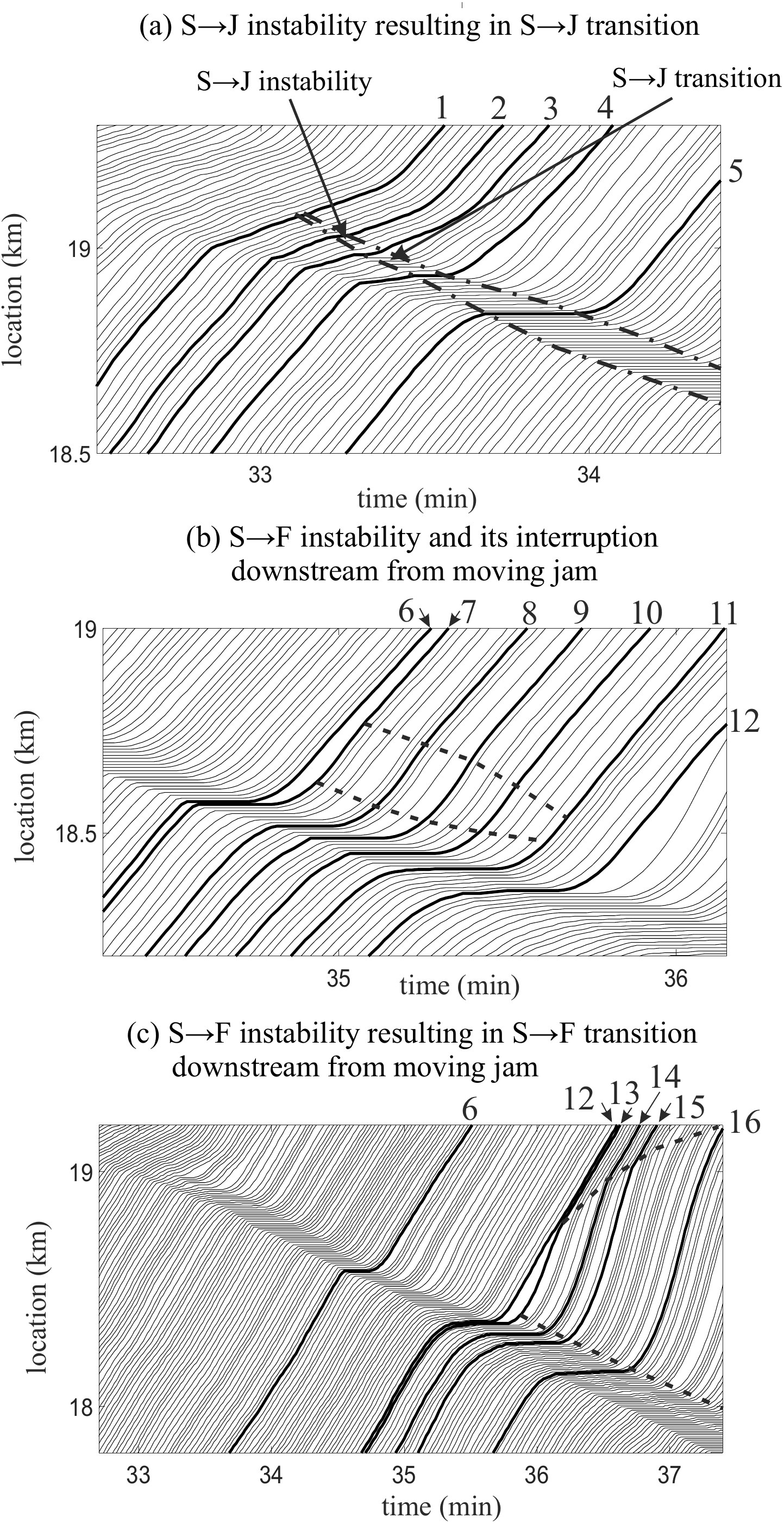}
\end{center}
\caption[]{Continuation of Fig.~\ref{Realizations_Fig} (b).
S$\rightarrow$J instability and resulting spatiotemporal nucleation traffic phenomena in synchronized flow:
(a) Vehicle trajectories  illustrating
S$\rightarrow$J instability resulting in S$\rightarrow$J transition.
(b) Vehicle trajectories for a later time interval as that in (a); illustration of
S$\rightarrow$F instability and its interruption   downstream   of     moving jam resulting
from the development of the S$\rightarrow$J instability shown in (a).
 (c) Vehicle trajectories for a longer time interval as that in (a); illustration of
S$\rightarrow$F instability that leads to S$\rightarrow$F transition   downstream   of    moving jam resulting
from the development of the S$\rightarrow$J instability shown in (a).
  }
\label{S_J_circle} 
\end{figure} 

  \begin{figure}
\begin{center}
\includegraphics*[width=8 cm]{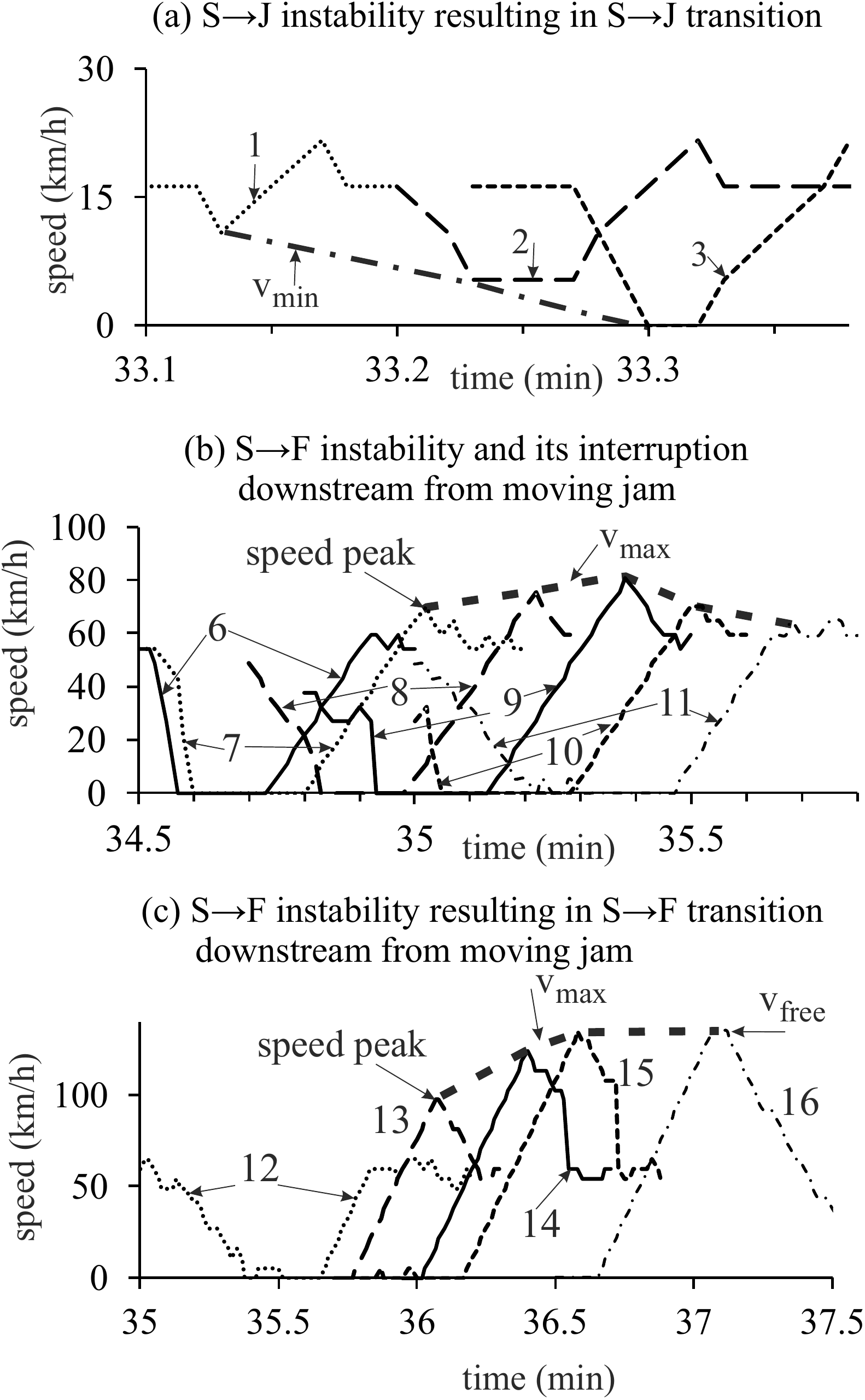}
\end{center}
\caption[]{Continuation of Fig.~\ref{S_J_circle}:
(a) Fragments of microscopic speeds related to Fig.~\ref{S_J_circle} (a); 
  bold dashed-dotted curve $v_{\rm min}(t)$
shows the time-dependence of the minimum speed on vehicle trajectories;  
  vehicle numbers in (a) are the same as those in   Fig.~\ref{S_J_circle} (a). 
(b) Fragments of microscopic speeds related to Fig.~\ref{S_J_circle} (b);
bold dashed curve $v_{\rm max}(t)$
shows the time-dependence of the maximum speed on vehicle trajectories;  
  vehicle numbers in (b) are the same as those in   Fig.~\ref{S_J_circle} (b).
 (c) Fragments of microscopic speeds related to Fig.~\ref{S_J_circle} (c);
bold dashed curve $v_{\rm max}(t)$
shows the time-dependence of the maximum speed on vehicle trajectories;  
  vehicle numbers in (c) are the same as those in   Fig.~\ref{S_J_circle} (c).
  }
\label{S_J_circle_speed} 
\end{figure}

Because the S$\rightarrow$J instability  shown in   Fig.~\ref{S_J_circle} (a) occurs in synchronized flow,
during the emergence of a moving jam, firstly, synchronized flow is realized
 both upstream and downstream of the moving jam
(trajectories 1 and 2 in Figs.~\ref{S_J_circle} (a) and~\ref{S_J_circle_speed} (a)).
Vehicles at the downstream jam front    start
their acceleration from the moving jam to   synchronized flow downstream of the moving jam
with a random time delay. Therefore,  time highway between two 
vehicles following each other while accelerating 
at the jam downstream front is a random value.

 Due to a random increase in time delay in vehicle acceleration, time headway
between two following vehicles accelerating at the jam downstream front
 can randomly become considerably longer than the mean time headway. For example, this case occurs between vehicles 6 and 7
shown in Fig.~\ref{S_J_circle} (b). In this case,
     following vehicle 7 accelerates to a higher speed
than   preceding vehicle 6. While approaching   
 vehicle 6,   vehicle 7 must decelerate to the synchronized flow speed of   vehicle 6.
As a result, a local speed increase
appears in synchronized flow just downstream of the moving jam that   we call {\it speed peak}
in synchronized flow downstream of moving jam
(labeled by $\lq\lq$speed peak" on vehicle trajectory 7 in Fig.~\ref{S_J_circle_speed} (b)).

It turns out that due to this speed peak  
  an S$\rightarrow$F instability can occur in synchronized flow downstream of the moving jam~\cite{SF_Nucleus}.
	An example of such S$\rightarrow$F instability initiated by the speed peak on vehicle trajectory 7 is shown in
	Figs.~\ref{S_J_circle} (b) and~\ref{S_J_circle_speed} (b): The maximum speed $v_{\rm max}$
	of following vehicles
	8 and 9 in Fig.~\ref{S_J_circle_speed} (b) increases.
	However, in the case under consideration, 
	the development of the S$\rightarrow$F instability is interrupted.
	Therefore, a dissolving speed wave of the local speed increase in synchronized flow
		is realized (wave labeled by dashed curves in Fig.~\ref{S_J_circle} (b)).
	This is qualitatively
	the same effect of the occurrence of a dissolving speed wave of the local speed increase in synchronized flow  as that shown in Fig.~\ref{S_J_F_circle} (a)
	(Sec.~\ref{Waves_Sub}). However in contract with Fig.~\ref{S_J_F_circle} (a), in the case, the dissolving speed wave of the local speed increase in synchronized flow occurs downstream of a moving jam
	(Figs.~\ref{S_J_circle} (b) and~\ref{S_J_circle_speed} (b)).

\subsubsection{S$\rightarrow$F instability resulting in S$\rightarrow$F transition
  downstream from  moving jam \label{S-J-res-S-F}}

However, in many other cases a speed peak occurring
due to a random increase in time delay in vehicle acceleration at the downstream front of a moving jam 
 can randomly become large enough to cause an S$\rightarrow$F transition. 
For example, such a long time delay  occurs between vehicles 12 and 13
shown in Figs.~\ref{S_J_circle} (c) and~\ref{S_J_circle_speed} (c).
As a result, a growing speed wave of the local speed increase in synchronized flow
		is realized (wave labeled by dashed curves in Fig.~\ref{S_J_circle} (c)).
In this case, 
	the development of an S$\rightarrow$F instability leads to an S$\rightarrow$F transition
	(speed of vehicle 16 reaches the maximum free flow speed $v_{\rm free}$ in Fig.~\ref{S_J_circle_speed} (c)).

	Thus, in this case the initial S$\rightarrow$J instability leads firstly to
	S$\rightarrow$J transition (labeled by $\lq\lq$S$\rightarrow$J transition"
	in Fig.~\ref{S_J_circle} (a)); later, due to a random time delay in vehicle acceleration
	at the downstream jam front, a speed peak occurs in synchronized flow downstream of
	the moving jam (Fig.~\ref{S_J_circle_speed} (c));
	the speed peak initiates 
 the S$\rightarrow$F instability; the development of the S$\rightarrow$F instability
 results in the S$\rightarrow$F transition
	(labeled by dashed curves
	in Figs.~\ref{S_J_circle} (c) and~\ref{S_J_circle_speed} (c)). In other words, the initial
	S$\rightarrow$J instability causes
	 a sequence of S$\rightarrow$J$\rightarrow$S$\rightarrow$F transitions.

  \begin{figure}
\begin{center}
\includegraphics*[width=8 cm]{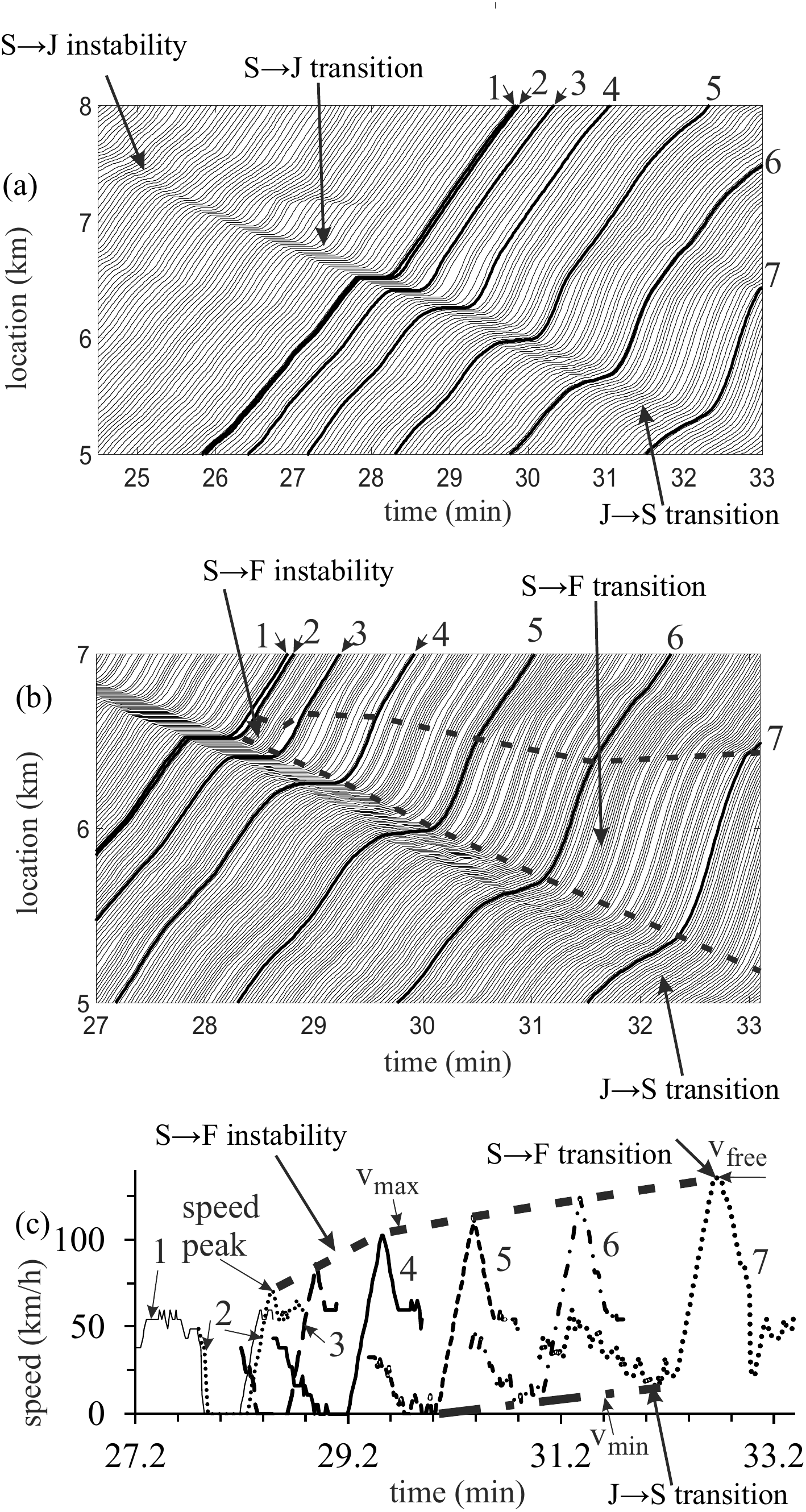}
\end{center}
\caption[]{Continuation of Fig.~\ref{Realizations_Fig} (c):
S$\rightarrow$J instability    resulting in S$\rightarrow$J transition with subsequent
S$\rightarrow$F transition   downstream of the moving jam
 following by jam dissolution (J$\rightarrow$S transition):
(a, b) Vehicle trajectories for different time intervals; in (a)
 each 2nd trajectory   and in (b) each trajectory   are shown.
(c) Fragments of microscopic speeds illustrating S$\rightarrow$F transition
  downstream of moving jam
	and  jam dissolution (J$\rightarrow$S transition); bold dashed and dashed-dotted
curves $v_{\rm max}(t)$ and $v_{\rm min}(t)$
show, respectively, the time-dependence of the maximum and minimum speeds on vehicle trajectories.  
  Vehicle numbers in (c) are the same as those in   (a, b).
  }
\label{S_J_circle_2} 
\end{figure}

There are simulation realizations in which
   sequences of S$\rightarrow$J$\rightarrow$S$\rightarrow$F transitions
 can exhibit some other features. An example is shown in Fig.~\ref{S_J_circle_2}.
Firstly, as in  Fig.~\ref{S_J_circle},
an initial S$\rightarrow$J instability leads  to
	S$\rightarrow$J transition (labeled by $\lq\lq$S$\rightarrow$J transition"
	in Fig.~\ref{S_J_circle_2} (a)). Then,   a speed peak occurs in synchronized flow downstream of
	the moving jam (Fig.~\ref{S_J_circle_2} (c)). The speed peak
	  initiates 
 the S$\rightarrow$F instability that development  
 results in the S$\rightarrow$F transition
	(Fig.~\ref{S_J_circle_2} (c)). Consequently, a
	   sequence of S$\rightarrow$J$\rightarrow$S$\rightarrow$F transitions is realized.
		However, in contrast with the 
		sequence of S$\rightarrow$J$\rightarrow$S$\rightarrow$F transitions shown in
		Figs.~\ref{S_J_circle} and~\ref{S_J_circle_speed}, in the case under consideration
		(Fig.~\ref{S_J_circle_2})
		due to the speed adaptation effect at the upstream jam front
		the wide moving jam dissolves over time and synchronized flow returns
		(labeled by $\lq\lq$J$\rightarrow$S transition" in 
		Figs.~\ref{S_J_circle_2} (a--c)).   

\subsection{S$\rightarrow$F instability leading to S$\rightarrow$F   transition \label{S-F-F}}

	An S$\rightarrow$F transition can also randomly occur in synchronized flow without  the effect
	of a moving jam discussed  in Sec.~\ref{S-J-res-S-F}.
	This case is shown
in simulation realization 4 (Fig.~\ref{Realizations_Fig} (d)):  
A large enough local speed increase in synchronized flow appears randomly initiating an S$\rightarrow$F instability.
The development of the S$\rightarrow$F instability is not interrupted over time; therefore,   the development of the S$\rightarrow$F instability leads to an S$\rightarrow$F transition. The microscopic development of this effect
is shown in   Fig.~\ref{S_F_circle}.

  \begin{figure}
\begin{center}
\includegraphics*[width=8 cm]{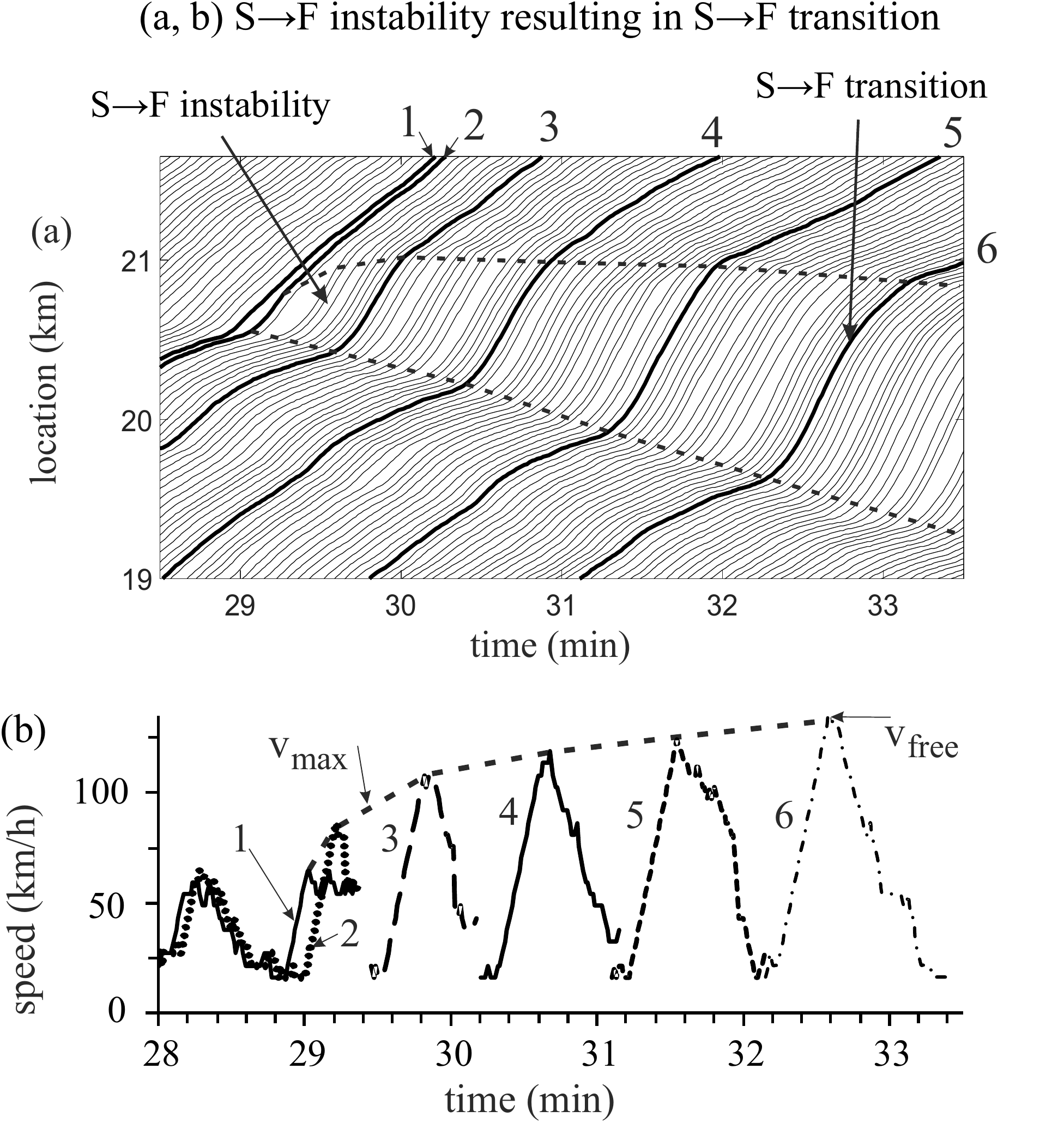}
\end{center}
\caption[]{Continuation of Fig.~\ref{Realizations_Fig} (d).
S$\rightarrow$F instability resulting in S$\rightarrow$F transition:
(a) Vehicle trajectories.
(b) Fragments of microscopic speeds illustrating
S$\rightarrow$F instability resulting in S$\rightarrow$F transition; bold dashed curve $v_{\rm max}(t)$
shows the time-dependence of the maximum speed on vehicle trajectories.  
  Vehicle numbers in (b) are the same as those in   (a).
  }
\label{S_F_circle} 
\end{figure} 

Within a road region of the development of the S$\rightarrow$F instability,
the vehicle speed increases subsequently over time. 
Because synchronized flow exists downstream of this local region, vehicles must decelerate approaching
slower moving vehicles ahead.
This results in a speed wave   within which
  the speed is larger than in surrounded synchronized flow (trajectories 2--4 in Fig.~\ref{S_F_circle}).
	The subsequent growth of this wave causes the S$\rightarrow$F transition 
	(trajectories 5, 6 in Fig.~\ref{S_F_circle}).
	
	\subsection{S$\rightarrow$F instability initiating S$\rightarrow$J instability:
S$\rightarrow$F$\rightarrow$S$\rightarrow$J transitions \label{S-F-ini-S-J}} 

 \begin{figure} 
\begin{center}
\includegraphics*[width=8 cm]{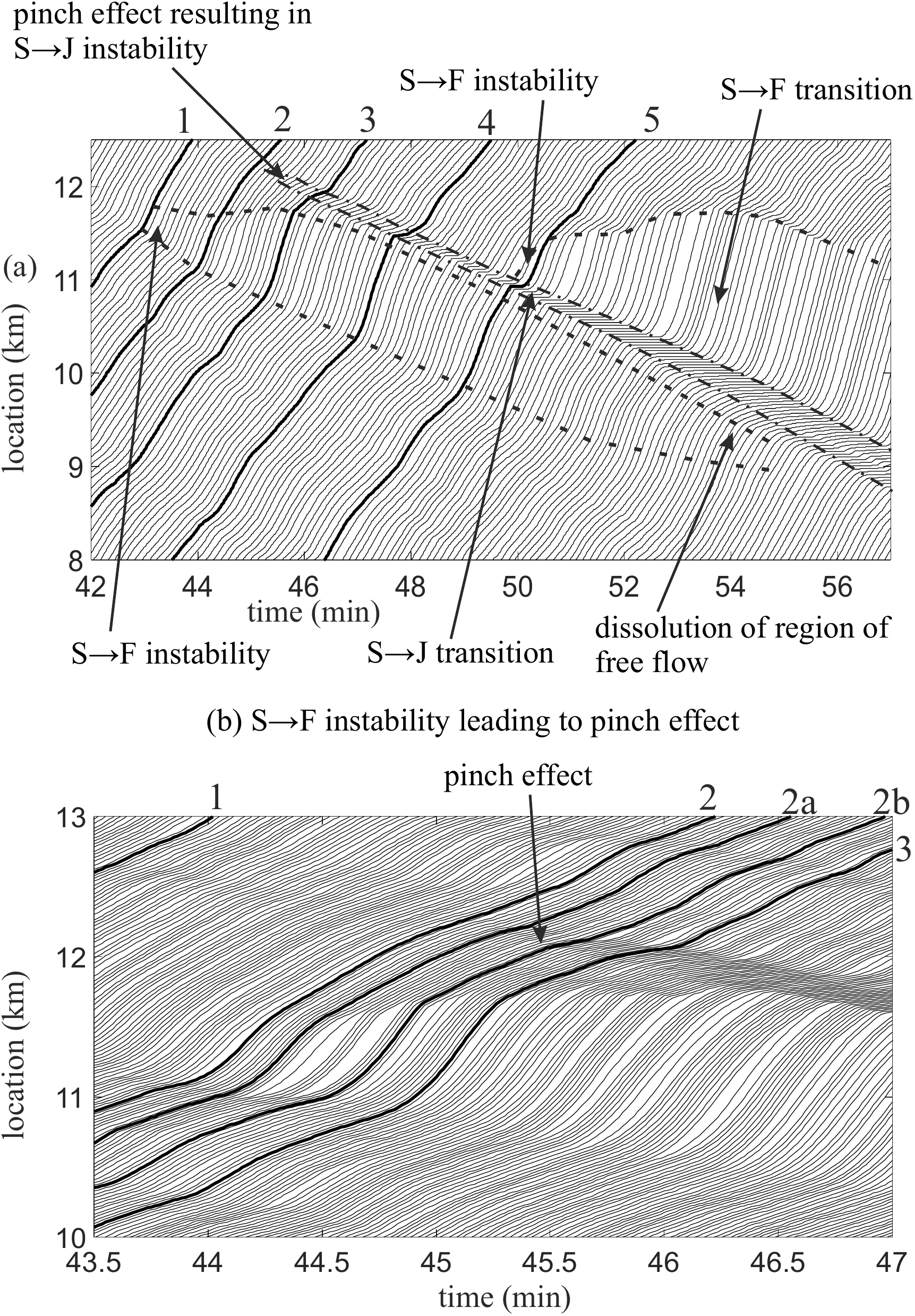}
\end{center}
\caption[]{Continuation of Fig.~\ref{Realizations_Fig} (e).
S$\rightarrow$F instability resulting in  S$\rightarrow$J  transition:
(a) Vehicle trajectories; each 4th vehicle trajectory is shown.
(b) Vehicle trajectories; each vehicle  trajectory is shown.
  }
\label{S_F_J_circle_pinch} 
\end{figure}

  \begin{figure}
\begin{center}
\includegraphics*[width=8 cm]{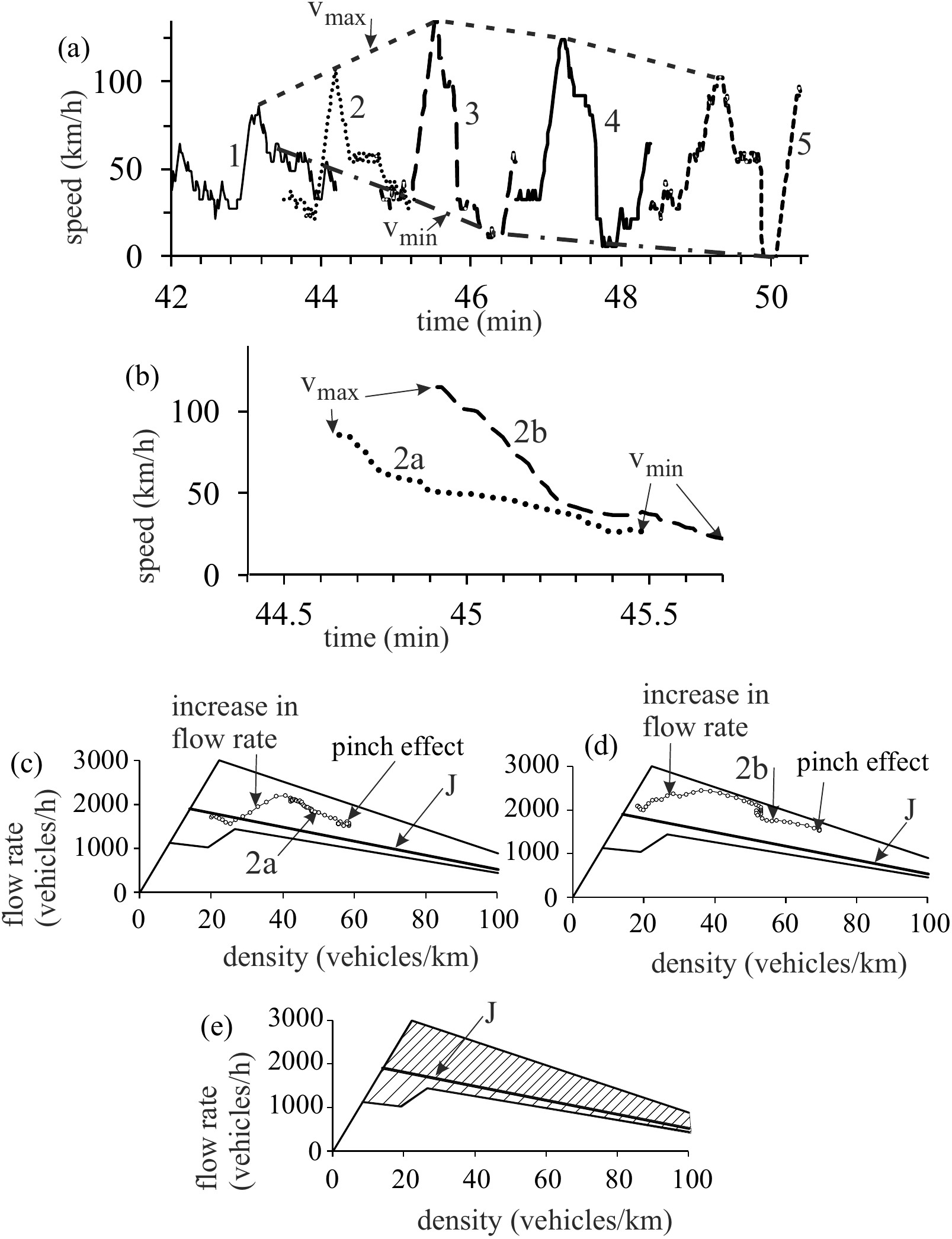}
\end{center}
\caption[]{Continuation of Fig.~\ref{S_F_J_circle_pinch}.
(a)  Fragments of microscopic speeds; 
  bold dashed and dashed-dotted curves $v_{\rm max}(t)$ and  $v_{\rm min}(t)$
show, respectively,  the time-dependence of the maximum and minimum speeds on vehicle trajectories;
vehicle numbers in (a) are the same as those in   Fig.~\ref{S_F_J_circle_pinch} (a).
(b) Fragments of microscopic speeds of vehicles shown by the same numbers in Fig.~\ref{S_F_J_circle_pinch} (b).
(c, d) Points along vehicle trajectories 2a (c) and 2b (d) in the flow--density plane;
averaging over 10 vehicles upstream of the related vehicle for each time step (1 sec). (e) Explanation of the pinch effect~\cite{KernerBook}:
Points on and above the line J are metastable with respect to S$\rightarrow$J transition; points below the line J
are stable with respect to S$\rightarrow$J instability.
  }
\label{S_F_J_circle_pinch_tr} 
\end{figure}

It turns out that when due to an S$\rightarrow$F instability  
  a growing speed wave of a local   speed increase is realized, then
due to a time delay in vehicle deceleration, vehicles decelerating at the downstream front of this growing
speed wave
 come on average at  small space gaps to each other.
As a result, the pinch effect can occur in synchronized flow that leads to
 an S$\rightarrow$J instability (emergence of a 
growing narrow moving jam(s)) resulting in an S$\rightarrow$J transition
(Figs.~\ref{S_F_J_circle_pinch} and~\ref{S_F_J_circle_pinch_tr}).
Thus, when the initial S$\rightarrow$F instability causes the pinch effect downstream
of the growing speed wave of a local speed increase in synchronized flow,
this S$\rightarrow$F instability
  can cause a sequence of phase transitions that can be called as
	S$\rightarrow$F$\rightarrow$S$\rightarrow$J transitions.

The development of the sequence of
 S$\rightarrow$F$\rightarrow$S$\rightarrow$J transitions is as follows.
The flow rate in synchronized flow   increases considerably just downstream of the
growing wave of the local speed increase caused by  the S$\rightarrow$F instability 
(labeled by $\lq\lq$increase in flow rate"
 on trajectories 2a and 2b in Figs.~\ref{S_F_J_circle_pinch_tr} (c, d)).
As known from other studies of dense synchronized flow
 (Fig.~\ref{S_F_J_circle_pinch_tr} (e))~\cite{KernerBook,Kerner2018B},
due to the flow rate increase in synchronized flow the pinch effect is often realized:
The density of synchronized flow
increases considerably and the speed decreases while   points related to the synchronized flow 
remain to be above the line $J$ in the flow-density plane 
(labeled by $\lq\lq$pinch effect" on points related to vehicles 2a and 2b in 
Figs.~\ref{S_F_J_circle_pinch_tr} (c, d)).
The narrow moving jam that has emerged in the pinch region of synchronized flow
transforms in a wide moving jam. Thus,
the initial  S$\rightarrow$F instability
 causes both the emergence of the local region of a higher speed and,
as the subsequent effect, the  S$\rightarrow$J instability resulting in the  S$\rightarrow$J transition.

In the example shown in Fig.~\ref{S_F_J_circle_pinch}, the growing speed wave of the local speed increase
(S$\rightarrow$F instability), which
  has led to the pinch effect resulting
in the  S$\rightarrow$J transition, transforms into a dissolving  speed wave of a local speed increase. Indeed,
the local region of free flow dissolves over time (labeled by $\lq\lq$dissolution of region of free flow"
in Fig.~\ref{S_F_J_circle_pinch} (a)). 

However, the pinch effect  
caused by the S$\rightarrow$F instability is often also observed
even when a local region of free flow resulting from
the initial S$\rightarrow$F instability does not dissolve over time.
An example is  a sequence of S$\rightarrow$F$\rightarrow$S$\rightarrow$J transitions
(labeled by $\lq\lq$S$\rightarrow$F$\rightarrow$S$\rightarrow$J" in Fig.~\ref{Realizations_Fig} (d)) 
caused by the S$\rightarrow$F instability discussed in Sec.~\ref{S-F-F}. In this 
sequence of S$\rightarrow$F$\rightarrow$S$\rightarrow$J transitions
  the resulting   wide moving jam is
labeled by $\lq\lq$jam" in Fig.~\ref{Realizations_Fig} (d).

\subsection{Independent development of both 
	S$\rightarrow$F and S$\rightarrow$J transitions in the same simulation realization}
	
	In Fig.~\ref{Realizations_Fig} (b--e), we have observed different   realizations
	simulated at the same model parameters. We have found that
	after  an   S$\rightarrow$F transition
	has occurred a subsequent (secondary) S$\rightarrow$J transition is possible (called as
	  S$\rightarrow$F$\rightarrow$S$\rightarrow$J transitions) (Sec.~\ref{S-F-ini-S-J});
		respectively, after  an   S$\rightarrow$J transition   
	has occurred a subsequent  (secondary) S$\rightarrow$F transition is possible (called as
	  S$\rightarrow$J$\rightarrow$S$\rightarrow$F transitions) (Sec.~\ref{S-J-S-F_trans}).
		In both cases, the secondary phase transition depends on the development of the initial one.
 However, simulations show that
 due to the competition between the S$\rightarrow$F and S$\rightarrow$J instabilities
	in initial synchronized flow, in other realizations  the independent development of 
	S$\rightarrow$F and S$\rightarrow$J transitions in different sections of the road
	at different random time instants is also possible 
	(Fig.~\ref{Realizations_Fig} (f, g)).
	
	In realization 6 (Fig.~\ref{Realizations_Fig} (f)), earlier
	an S$\rightarrow$J instability has occurred. As a result, a region (labeled by  bracket I)
	in which  complex spatiotemporal alternations of wide moving jams, free flow, and synchronized flow 
	are formed occurs; later in the   synchronized flow outside this region, due to
an S$\rightarrow$F instability  other alternations of traffic phases J, S, and F
are formed (labeled by  bracket II in Fig.~\ref{Realizations_Fig} (f)). Similar phenomena are observed in
realization 7 (Fig.~\ref{Realizations_Fig} (g)). In this case, earlier through the development of
an S$\rightarrow$F instability alternations of traffic phases J, S, and F
are formed (labeled by  bracket I in Fig.~\ref{Realizations_Fig} (g)); later another region 
of  alternations of traffic phases J, S, and F
is realized randomly (labeled by  bracket II).

\section{Probabilistic features of competition between
	S$\rightarrow$F and S$\rightarrow$J instabilities at highway bottleneck  \label{Probabilistic_On_S}}  
	
	In real traffic, synchronized flow occurs usually at bottlenecks. In this section, we show that the   statistical physics of
	synchronized flow presented in Sec.~\ref{Probabilistic_S}
	can   explain   statistical features of
	synchronized flow at road bottlenecks. For simplicity, we consider synchronized flow
 at an on-ramp bottleneck on a single-lane road. Our simulations   have shown that results of this analysis remain qualitatively for other bottleneck types.
	It is known that synchronized flow at the bottleneck occurs due to a spontaneous F$\rightarrow$S transition
	that exhibits  a random  time delay~\cite{KernerBook}. The physics of a random time delay is explained by a sequence of random F$\rightarrow$S$\rightarrow$F transitions
	at the bottleneck~\cite{Kerner2015B}. Because the F$\rightarrow$S$\rightarrow$F transitions have already been studied~\cite{Kerner2015B}, in simulations rather than a spontaneous F$\rightarrow$S transition,
	synchronized flow at the bottleneck that statistical features should be studied
	results from an induced F$\rightarrow$S transition
	(Fig.~\ref{Realizations_on}).
	To induce the F$\rightarrow$S transition, in all simulation realizations
	presented in Fig.~\ref{Realizations_on} we apply the same
	on-ramp inflow impulse of the amplitude
	$q^{\rm (imp)}_{\rm on}=$ 795 vehicles/h within   time interval $0\leq t\leq$ 4 min;
	 at $t>$ 4 min
	a given on-ramp inflow rate $q_{\rm on}$   is time-independent.
	
	\subsection{Probabilities of  
	S$\rightarrow$F and S$\rightarrow$J transition at bottleneck as functions of  flow rate}
	
\begin{figure}
\begin{center}
\includegraphics*[width=8 cm]{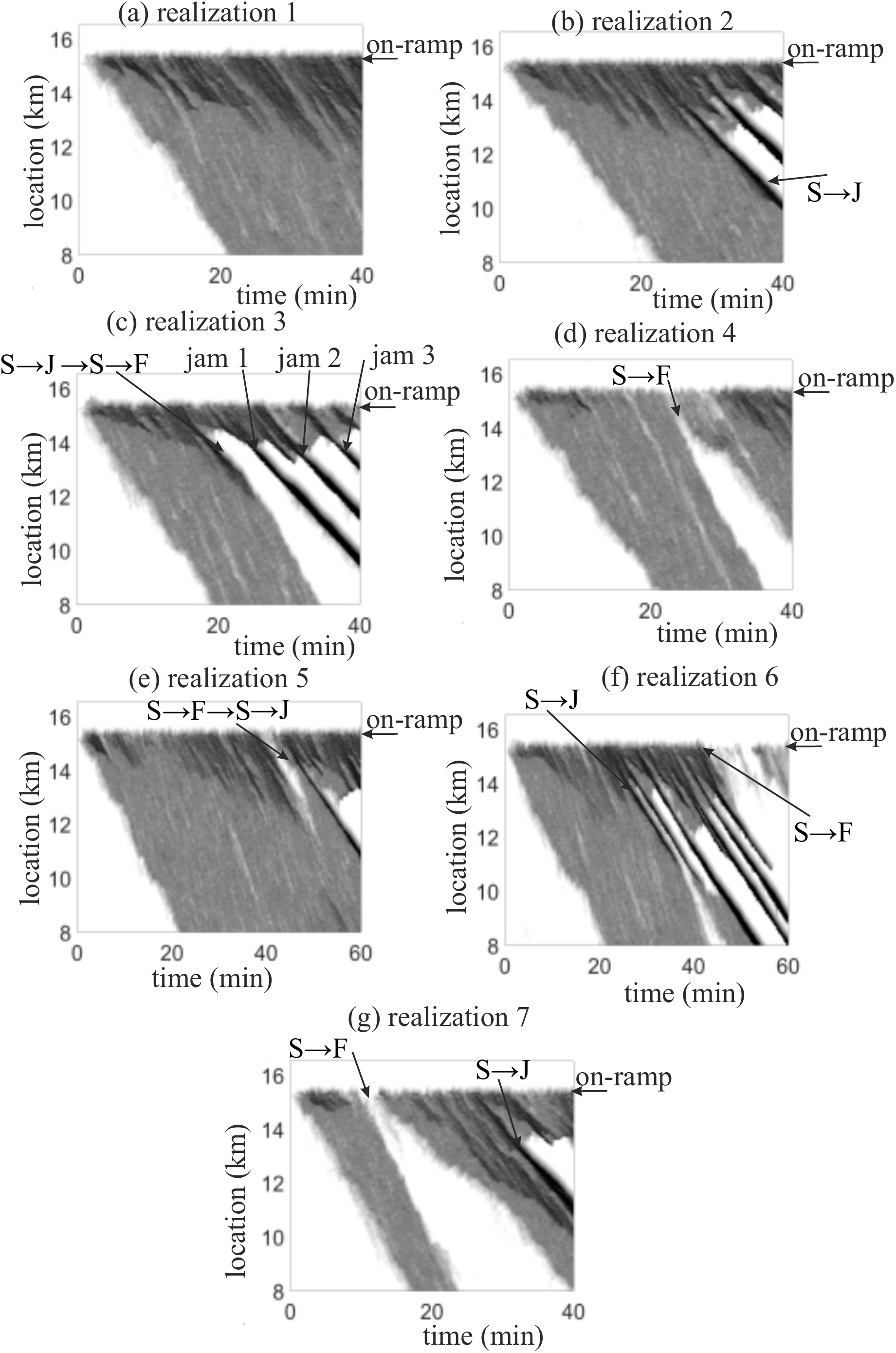}
\end{center}
\caption[]{Characteristic simulation realizations of synchronized flow at on-ramp bottleneck
 calculated  at the same 
	set of   parameters
  of the KKSW CA model, parameters of on-ramp bottleneck, and of the
	flow rates~\cite{Rand}: Vehicle  speed data presented by  regions with 
variable shades of gray    (shades of gray vary from white to black when the speed 
decreases from 120 km/h (white) to 0 km/h (black)).   $q_{\rm in}=$ 1914 vehicles/h.   $q_{\rm on}=$ 374 vehicles/h.
	 The beginning and   end of the on-ramp merging region are, respectively,
	$x_{\rm on}=$ 15 and $x^{\rm (e)}_{\rm on}=$ 15.3 km. 
	Arrows S$\rightarrow$F label  S$\rightarrow$F transition, arrows S$\rightarrow$J label S$\rightarrow$J transition.
  }
\label{Realizations_on} 
\end{figure}

	At the same model parameters and the same values of the flow rate in free flow upstream of the bottleneck $q_{\rm in}$ and the on-ramp inflow rate  $q_{\rm on}$,
	there are a   number of simulation realizations
in which different spatiotemporal critical traffic phenomena in synchronized flow at the bottleneck 
have been found. Characteristic simulation realizations
	are shown in Fig.~\ref{Realizations_on}.   
	
We have found that due to the  spatiotemporal competition
between
	S$\rightarrow$F and S$\rightarrow$J instabilities in synchronized   flow there is a range of 	  the flow rate   $q_{\rm sum}$ 
within which 
 during the observation time $T_{\rm ob}$  either synchronized flow persists
 (Fig.~\ref{Realizations_on} (a)), or firstly an S$\rightarrow$J transition occurs in synchronized flow
 (Fig.~\ref{Realizations_on} (b, c, f)), or else firstly an S$\rightarrow$F transition occurs in synchronized flow
 (Fig.~\ref{Realizations_on} (d, e, g)) with different probabilities, respectively, $P_{\rm S}(q_{\rm sum})$,
$P_{\rm SJ}(q_{\rm sum})$, and $P_{\rm SF}(q_{\rm sum})$.   Here and below the flow rate $q_{\rm sum}$ is equal to
\begin{equation}
q_{\rm sum}=q_{\rm on}+q_{\rm in}.
\label{q_sum_F}
\end{equation}

\begin{figure}
\begin{center}
\includegraphics*[width=7.5 cm]{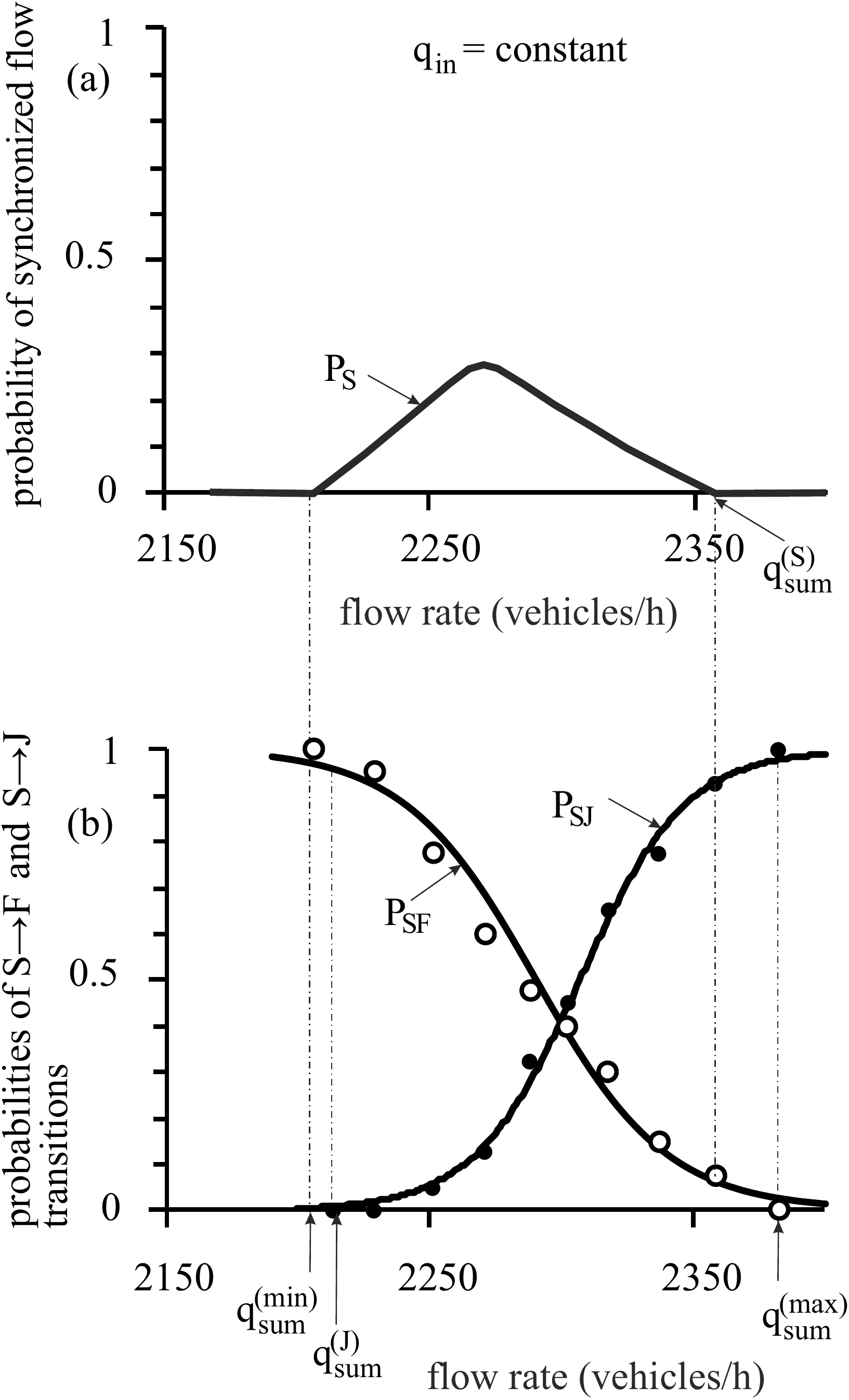}
\end{center}
\caption[]{Probabilities 
   $P_{\rm S}(q_{\rm sum})$ (curve $P_{\rm S}$) (a) and
	$P_{\rm SF}(q_{\rm sum})$ (curve $P_{\rm SF}$),
	$P_{\rm SJ}(q_{\rm sum})$ (curve $P_{\rm SJ}$) (b)
as functions  of the flow rate $q_{\rm sum}=q_{\rm on}+q_{\rm in}$
calculated through the change in   $q_{\rm on}$
at given flow rate $q_{\rm in}=$ 1914 vehicles/h.
For calculation of   $P_{\rm SF}(q_{\rm sum})$ and $P_{\rm SJ}(q_{\rm sum})$,
at each given value of $q_{\rm sum}$
	different simulation realizations (runs) $N_{\rm r}=$ 40 during the   time interval  
	$T_{\rm ob}=$ 40 min at the same set of model parameters~\cite{Rand} have been calculated.
  $P_{\rm SF}(q_{\rm sum})=n^{\rm (SF)}_{\rm r}/N_{\rm r}$,
	$P_{\rm SJ}(q_{\rm sum})=n^{\rm (SJ)}_{\rm r}/N_{\rm r}$, where $n^{\rm (SF)}_{\rm r}$ is the number of realizations in which S$\rightarrow$F transition  has firstly occurred during the time interval $T_{\rm ob}$,
	$n^{\rm (SJ)}_{\rm r}$ is the number of realizations in which S$\rightarrow$J transition  has firstly occurred during the time interval $T_{\rm ob}$;
	respectively, $P_{\rm S}(q_{\rm sum})=1-(P_{\rm SF}(q_{\rm sum})+P_{\rm SJ}(q_{\rm sum}))$.
	Other  model parameters are the same as those in Fig.~\ref{Realizations_on}.
	Calculated values: $q^{\rm (min)}_{\rm sum}=$ 2206, $q^{\rm (J)}_{\rm sum}=$
	2213, $q^{\rm (S)}_{\rm sum}=$ 2358, $q^{\rm (max)}_{\rm sum}=$  2382 
	     vehicles/h;
	$\alpha=$0.04 h/vehicles and $q_{\rm 0}=$ 2290 vehicles/h in (\ref{Prob_SF_on}),
	$\alpha=$0.05 h/vehicles and $q_{\rm 0}=$ 2307 vehicles/h in (\ref{Prob_SJ_on}).
  }
\label{Prob_on} 
\end{figure}

It has been found that the flow rate dependencies of these probabilities $P_{\rm S}(q_{\rm sum})$,
$P_{\rm SJ}(q_{\rm sum})$, and $P_{\rm SF}(q_{\rm sum})$ (Fig.~\ref{Prob_on}) are qualitatively the same as those for a homogeneous road (Fig.~\ref{Prob_Fig}) found in Sec.~\ref{Probabilistic_S}. Respectively, if 
the average space gap $\overline g$ in formulas (\ref{Prob2_F_gen})--(\ref{PP_F}) is replaced by the  flow rate $q_{\rm sum}$, then we come to formula $P_{\rm S}(q_{\rm sum})+P_{\rm SF}(q_{\rm sum})+P_{\rm SJ}(q_{\rm sum})=1$ as well as 
formulas
\begin{equation}
P_{\rm S}(q_{\rm sum})>0 \quad {\rm only \ if} \  q^{\rm (min)}_{\rm sum}  <  q_{\rm sum} <  q^{\rm (S)}_{\rm sum},
\label{Prob2_on}
\end{equation}
\begin{equation}
P_{\rm S}(q_{\rm sum})=0 \quad {\rm  if} \  q_{\rm sum} \leq q^{\rm (min)}_{\rm sum} \ {\rm or} \
 q_{\rm sum} \geq q^{\rm (S)}_{\rm sum},
\label{Prob3_on}
\end{equation}
\begin{equation}
P_{\rm SF}(q_{\rm sum})>0 \ {\rm and} \  P_{\rm SJ}(q_{\rm sum})>0 \ {\rm at} \   
q^{\rm (J)}_{\rm sum} < q_{\rm sum}    < q^{\rm (max)}_{\rm sum}.
\label{PP_on}
\end{equation}
Formulas (\ref{Prob2_on})--(\ref{PP_on})  determine some characteristic 
 flow rates $q^{\rm (S)}_{\rm sum}$, $q^{\rm (min)}_{\rm sum}$,
$q^{\rm (max)}_{\rm sum}$, and $q^{\rm (J)}_{\rm sum}$ (Fig.~\ref{Prob_on}).
These characteristic flow rates have qualitatively the same sense 
as the characteristic average space gaps in synchronized flow on  a homogeneous road in formulas   (\ref{Prob2_F})--(\ref{PP_F})
 (Sec.~\ref{Probabilistic_S}).

The flow-rate function of the   probability
$P_{\rm SF}(q_{\rm sum})$ (empty circles in Fig.~\ref{Prob_on} (b))
      is well  fitted by a function:
\begin{equation}
P_{\rm SF}(q_{\rm sum})=\frac{1}{1+ {\rm exp}[\alpha(q_{\rm sum}-q_{\rm 0})]};
\label{Prob_SF_on}
\end{equation}
 respectively,
the flow-rate function of the   probability
$P_{\rm SJ}(q_{\rm sum})$ (black circles in Fig.~\ref{Prob_on} (b))
      is well  fitted by a function:
\begin{equation}
P_{\rm SJ}(q_{\rm sum})=\frac{1}{1+ {\rm exp}[-\alpha(q_{\rm sum}-q_{\rm 0})]},
\label{Prob_SJ_on}
\end{equation}
where $q_{\rm sum}$ is the flow rate at the
 bottleneck (\ref{q_sum_F}), $\alpha$ and $q_{\rm 0}$ are constants (Fig.~\ref{Prob_on}).

As we can see, probabilistic features of synchronized flow at the bottleneck (Fig.~\ref{Prob_on}) are qualitatively
the same as those on a homogeneous road (Fig.~\ref{Prob_Fig}). However, there is a basic difference
between these two cases: On a homogeneous road, critical local disturbances initiating
the S$\rightarrow$F and S$\rightarrow$J instabilities
appear  at random road locations. On contrarily, as well-known the bottleneck introduces a large local inhomogeneity in traffic. For this reason, one can expect that critical local disturbances should 
appear randomly mostly in a vicinity of the bottleneck. Although this   conclusion
is obvious from earlier studies of traffic at a bottleneck~\cite{KernerBook,Kerner2015B}, as we will see below,
  due to the competition of
	S$\rightarrow$F and S$\rightarrow$J instabilities  a number of unknown before
	spatiotemporal traffic phenomena can be found at a bottleneck.

\subsection{Effect of  competition of
	S$\rightarrow$F and S$\rightarrow$J instabilities on  speed waves in
	synchronized flow at bottleneck \label{On_Dis_Speed_Waves}}

Speed (and, respectively, density) waves in synchronized flow
caused either by the S$\rightarrow$J instability~\cite{KernerBook}
or by the S$\rightarrow$F instability~\cite{Kerner2015B}
are known.   
It has been unknown that due to 
	the competition of
	S$\rightarrow$F and S$\rightarrow$J instabilities at the  bottleneck,
	in the same realization 
random alternations of these both instabilities  can occur
leading to 	speed and density waves
in synchronized flow that are non-regular in space and time
 (Fig.~\ref{On_Waves})~\cite{Wave_MSPs}.

		  \begin{figure}
\begin{center}
\includegraphics*[width=8 cm]{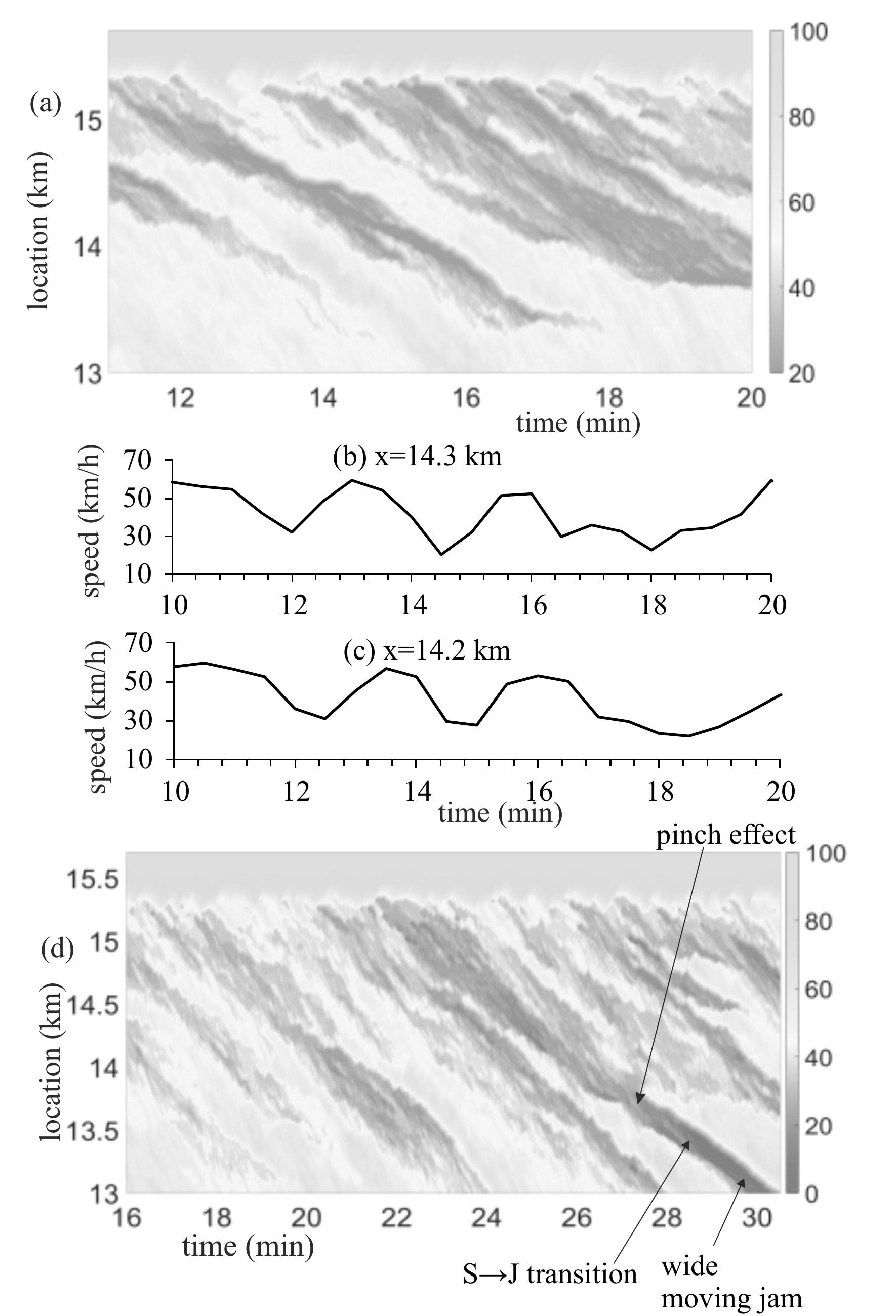}
\end{center}
\caption[]{Waves in synchronized flow at bottleneck:
(a) Fragment of vehicle  speed data presented by  regions with 
variable shades of gray   related to Fig.~\ref{Realizations_on} (a).
(b, c) Time-functions of average speed
 (30 sec averaged data) measured by virtual detectors at two different locations in (a).
(d) Fragment of vehicle data presented by  regions with 
variable shades of gray  related to Fig.~\ref{Realizations_on} (b). 
In (a, d), the beginning and   end of the on-ramp merging region are, respectively,
	$x_{\rm on}=$ 15 and $x^{\rm (e)}_{\rm on}=$ 15.3 km. 
  }
\label{On_Waves} 
\end{figure}

	In particular, these waves occur through the effect of
	an S$\rightarrow$F instability and its interruption
	as well as the effect an S$\rightarrow$J instability and its interruption
	occurring randomly at the bottleneck at different  time instants
	as shown, respectively,
	in Fig.~\ref{On_Waves2} (a, b) and
	Fig.~\ref{On_Waves2} (c, d).
To find the reason for the alternation of these two different effects,
we should consider initial local speed disturbances occurring due to vehicle merging from the on-ramp
onto the main road within the on-ramp merging region (15.3 km $\leq x \leq$ 15 km) at the bottleneck
(Fig.~\ref{On_Waves2} (a, c)). There are two qualitatively different kinds of
the initial local disturbances: (i)  
$\lq\lq$Speed peak" (Fig.~\ref{On_Waves2} (b)) occurs due to the merging of vehicle 2 from on-ramp
(dotted curves 2 in Figs.~\ref{On_Waves2} (a, b)):
The motion of downstream  vehicle 1 is not influenced by vehicle 2. It is different for
 upstream   vehicle 3, which has earlier accelerated while following vehicle 1; after the merging of vehicle 2,
vehicle 3 must interrupt its acceleration and decelerate  strongly to avoid the 
collision with vehicle 2~\cite{Kerner2015B}.  
 (ii) $\lq\lq$Local speed reduction"
(Fig.~\ref{On_Waves2} (d)) occurs also due to the merging of vehicle 9 from on-ramp
(dotted curves 9 in Figs.~\ref{On_Waves2} (c, d);  in this case, in contrast
with the speed peak, upstream   vehicle 10 has not still earlier accelerated behind vehicle 8, therefore, vehicle 10, while decelerating behind slow moving vehicle 9, produces no speed peak~\cite{KernerBook}.

		  \begin{figure}
\begin{center}
\includegraphics*[width=8 cm]{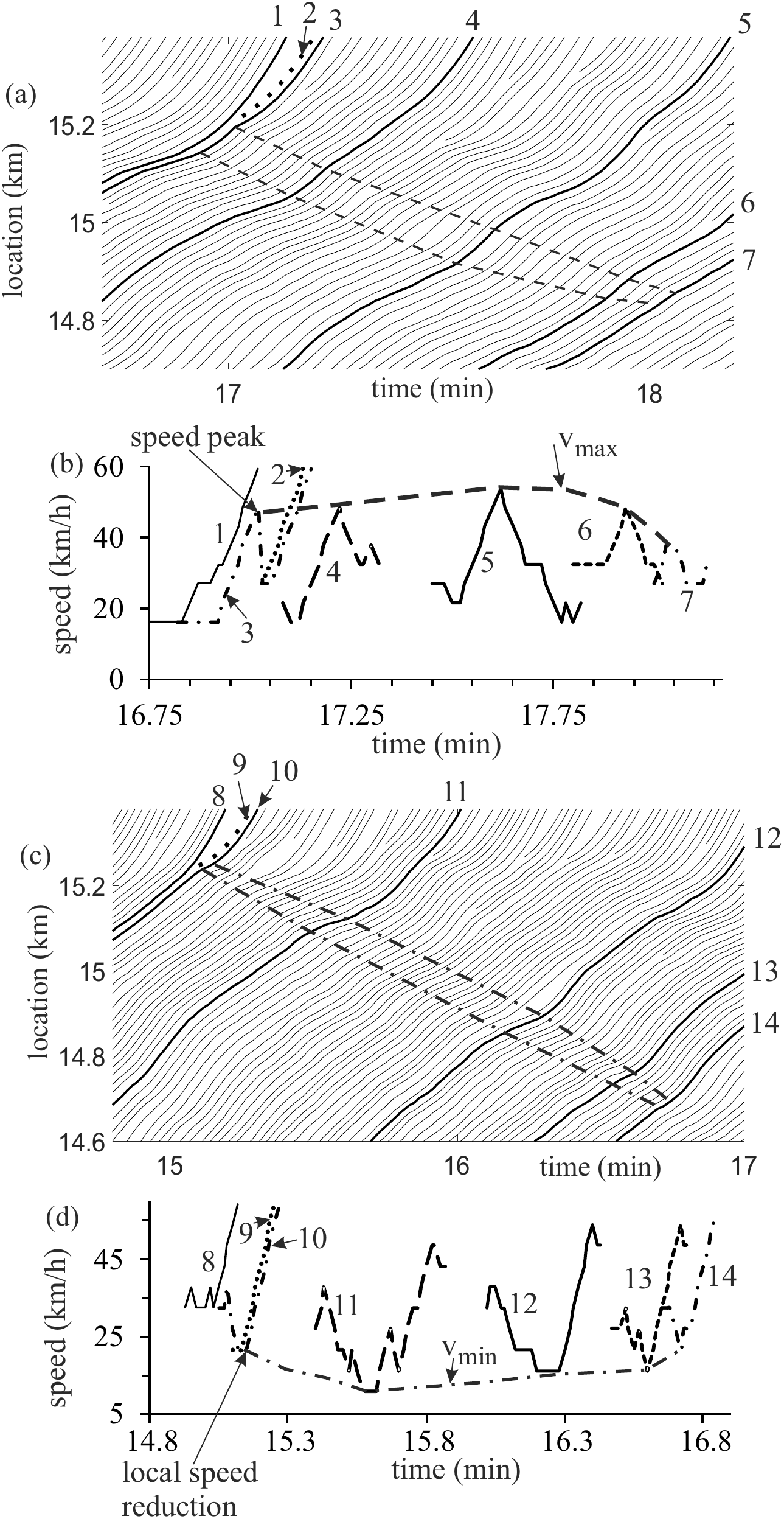}
\end{center}
\caption[]{Continuation of Fig.~\ref{On_Waves} (a):
(a, b) S$\rightarrow$F instability and its interruption.
(c, d) S$\rightarrow$J instability and its interruption.
(a, c) Fragments of vehicle trajectories.
(b, d) Fragments of time-functions of microscopic speeds for different vehicles
 whose numbers are the same as those in (a, c), respectively. 
In (a, c), the beginning and   end of the on-ramp merging region are, respectively,
	$x_{\rm on}=$ 15 and $x^{\rm (e)}_{\rm on}=$ 15.3 km. 
  }
\label{On_Waves2} 
\end{figure}

These different disturbances at the bottleneck cause
two different effects: 1. The speed peak initiates 
an S$\rightarrow$F instability: A speed wave of local speed increase
grows while propagating upstream (trajectories 3--5 in Fig.~\ref{On_Waves2} (b)); however,
in the case under consideration 
the development of the S$\rightarrow$F instability is interrupted (trajectories  5--7
 in Fig.~\ref{On_Waves2} (b)) and, therefore, no S$\rightarrow$F transition occurs. 2. 
The local speed reduction initiates 
an S$\rightarrow$J instability: A speed wave of local speed decrease (narrow moving jam)
grows while propagating upstream (trajectories 10 and 11 in Fig.~\ref{On_Waves2} (d)); however,
in the case under consideration 
the development of the S$\rightarrow$J instability is interrupted (trajectories  11--14
 in Fig.~\ref{On_Waves2} (d)) and, therefore, no wide moving jam (S$\rightarrow$J transition) occurs.
In Fig.~\ref{On_Waves} (a), all speed waves dissolve
 over time (dissolving speed waves), i.e.,
neither 
S$\rightarrow$F  transition nor S$\rightarrow$J transition is realized. 

In some realizations, within a dissolving wave of the local speed increase in synchronized flow
  randomly a  local increase in the speed can appear at some distance upstream of the bottleneck causing an 
S$\rightarrow$F  transition. Such   cases associated with 
  the dissolving wave found in simulations are not shown.
This is because such cases are relatively seldom. In the other realizations, in which
  a speed peak appears
  just  
	at the bottleneck location while initiating
  an S$\rightarrow$F  instability, 
the development of the S$\rightarrow$F instability leads to an S$\rightarrow$F transition
 (Figs.~\ref{Realizations_on} (d, e, f, g))
	as already studied in~\cite{Kerner2015B}.
	However,  as found in this paper rather than
	the latter known scenario of the development of
	the S$\rightarrow$F transition~\cite{Kerner2015B},
	many of the  S$\rightarrow$F transitions that occur upstream of the bottleneck 
	result from a qualitatively different scenario:
	These S$\rightarrow$F transitions are the consequence of
	initial	 S$\rightarrow$J instabilities leading to
	 sequences of S$\rightarrow$J$\rightarrow$S$\rightarrow$F transitions
 (see Sec.~\ref{On_S-J-S_on}).

In some other realizations, within a   wave of the local speed decrease in synchronized flow
  randomly the pinch effect  is realized  at 
	a considerable distance upstream of the
	bottleneck. Such a  case of the pinch effect with the subsequent  S$\rightarrow$J  transition is shown in Fig.~\ref{On_Waves} (d).  
In other realizations,   
an S$\rightarrow$J instability (pinch effect with growing narrow moving jams) occurs   just at the bottleneck
location 
(Figs.~\ref{Realizations_on} (c, f, g)); in these cases,   the S$\rightarrow$J instability leads to an S$\rightarrow$J transition
	as already studied in~\cite{KernerBook}.
However, as found in this paper  rather than
	the latter well-known scenario of the development of
	the S$\rightarrow$J transition~\cite{KernerBook},    many of the
	   S$\rightarrow$J transitions that occur upstream of the bottleneck result from a qualitatively different scenario:
		These S$\rightarrow$J transitions are the consequence of
	initial	 S$\rightarrow$F instabilities leading to
	 sequences of S$\rightarrow$F$\rightarrow$S$\rightarrow$J transitions 
	  (see Sec.~\ref{On_S-F-J_on}).

\subsection{S$\rightarrow$J$\rightarrow$S$\rightarrow$F transitions  at bottleneck \label{On_S-J-S_on}}
 
Sequences of S$\rightarrow$J$\rightarrow$S$\rightarrow$F transitions
found for 
  a homogeneous road (Sec.~\ref{S-J-S-F_trans}) are often observed in synchronized flow
at the bottleneck.

\begin{figure}
\begin{center}
\includegraphics*[width=8 cm]{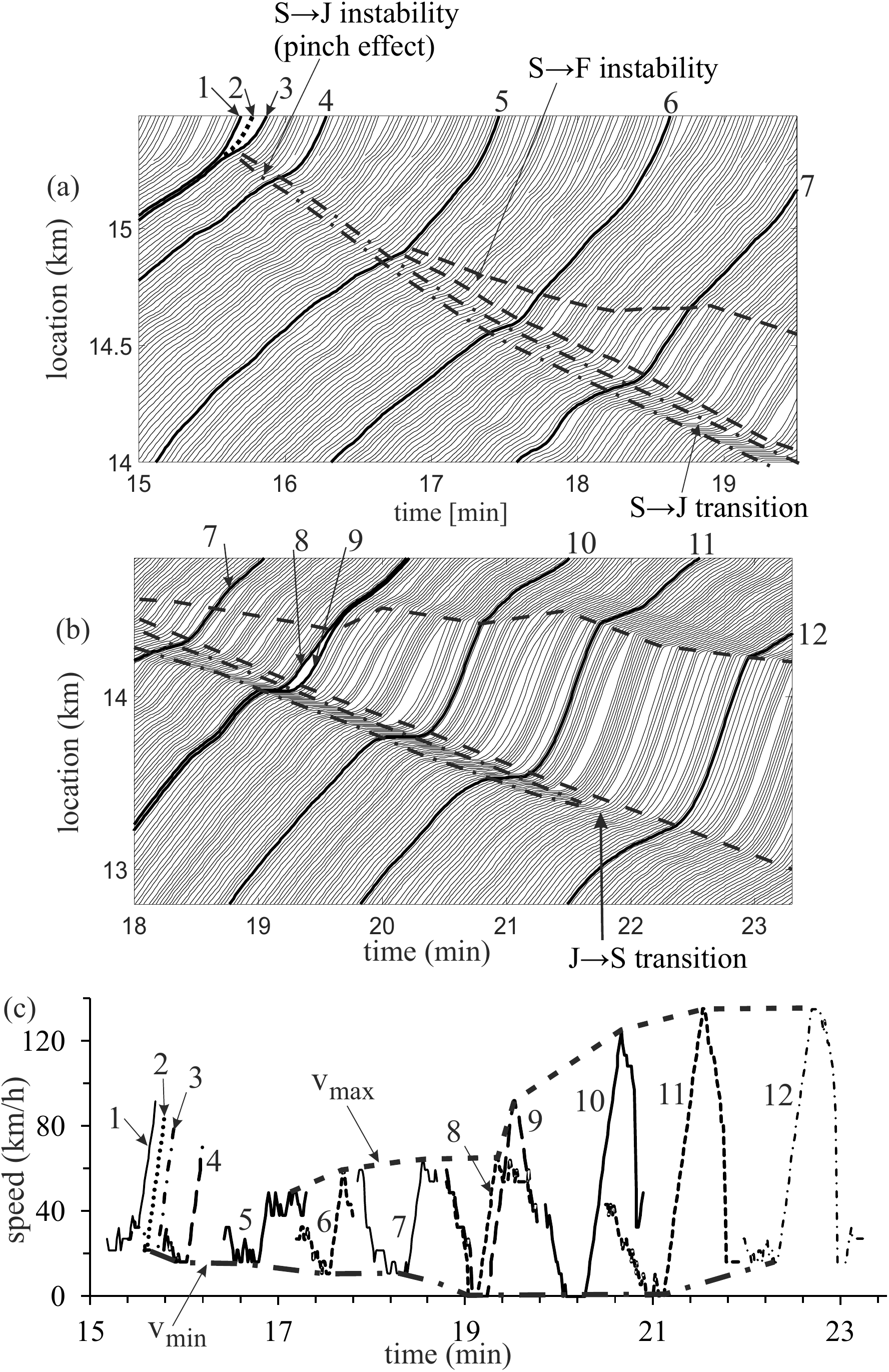}
\end{center}
\caption[]{Continuation of Fig.~\ref{Realizations_on} (c).
S$\rightarrow$J$\rightarrow$S$\rightarrow$F transitions  at bottleneck:
(a, b) Fragments of vehicle trajectories.
(c) Fragments of microscopic speeds; 
  bold dashed and dashed-dotted curves $v_{\rm max}(t)$ and  $v_{\rm min}(t)$
show, respectively,  the time-dependence of the maximum and minimum speeds on vehicle trajectories.
Vehicle numbers in (c) are the same as those in (a, b).
In (a, b), the beginning and   end of the on-ramp merging region are, respectively,
	$x_{\rm on}=$ 15 and $x^{\rm (e)}_{\rm on}=$ 15.3 km.  
  }
\label{On_J_S_F} 
\end{figure}

An example of S$\rightarrow$J$\rightarrow$S$\rightarrow$F transitions in synchronized flow at the bottleneck
is presented in  Fig.~\ref{On_J_S_F}.
As for the homogeneous road (Fig.~\ref{S_J_circle_2}), a moving jam
that emerges due to the development of
an S$\rightarrow$J instability
 in synchronized flow at the bottleneck   results in an
 S$\rightarrow$F instability with the subsequent an S$\rightarrow$F transition
   downstream of the moving jam (Fig.~\ref{On_J_S_F})~\cite{Concave_S}.
 As on homogeneous road (Fig.~\ref{S_J_circle_2}),  
the wide moving jam dissolves over time (Fig.~\ref{On_J_S_F} (b, c)). 
However, it should be emphasized that
such a jam dissolution (Fig.~\ref{On_J_S_F} (b, c))
 is not a general case:
In other cases  of the  occurrence of a sequence of
S$\rightarrow$J$\rightarrow$S$\rightarrow$F transitions at the bottleneck, both
the wide moving jam and free flow   downstream of the moving jam persist over time.
This is realized even in the same  realization 3 (persisting wide moving jams
are labeled by $\lq\lq$jam 1", $\lq\lq$jam 2", and $\lq\lq$jam 3" in Fig.~\ref{Realizations_on} (c)).

The   similarity of
S$\rightarrow$J$\rightarrow$S$\rightarrow$F transitions at the bottleneck and on the homogeneous road
is explained as follows: After a moving jam  
has occurred in synchronized flow at the bottleneck, the subsequent development
of  an S$\rightarrow$F instability downstream of the moving jam occurs
 far  upstream of the bottleneck. Therefore, this S$\rightarrow$F instability
does not  almost depend on the bottleneck:
The bottleneck is the reason for synchronized flow occurrence.
After synchronized flow has already occurred at the bottleneck, further critical phenomena in synchronized flow
like S$\rightarrow$J$\rightarrow$S$\rightarrow$F transitions are qualitatively the same for a hypothetical homogeneous   road and for real roads with bottlenecks.

\subsection{S$\rightarrow$F$\rightarrow$S$\rightarrow$J transitions   at bottleneck \label{On_S-F-J_on}}

Sequences of S$\rightarrow$F$\rightarrow$S$\rightarrow$J transitions
found for 
  a homogeneous road (Sec.~\ref{S-F-ini-S-J}) is also a characteristic effect in synchronized flow
at the bottleneck (Figs.~\ref{On_S_F_J}
and~\ref{On_S_F_J_pinch_tr}) caused by the competition of the S$\rightarrow$F and S$\rightarrow$J instabilities.

			  \begin{figure}
\begin{center}
\includegraphics*[width=8 cm]{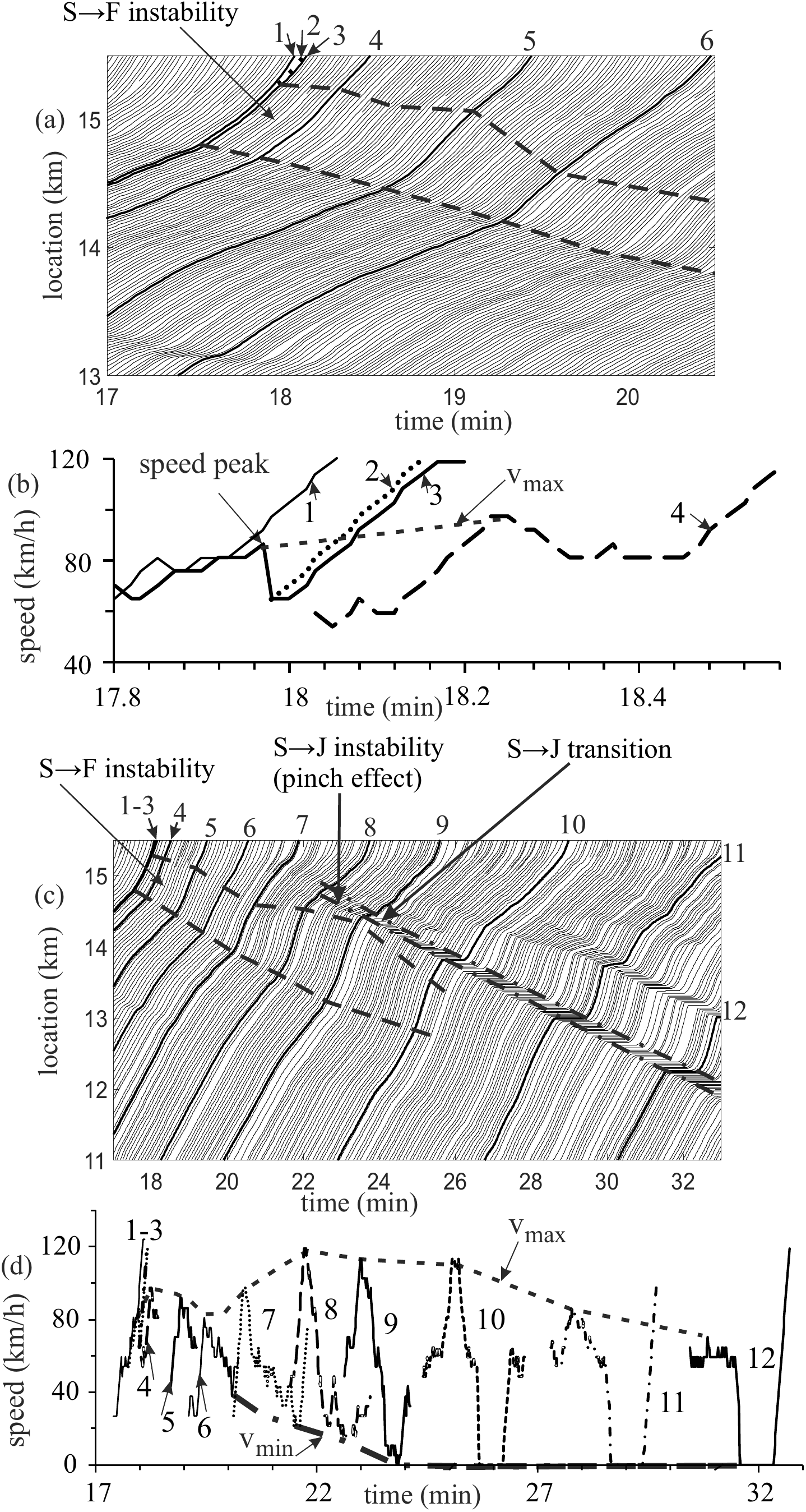}
\end{center}
\caption[]{Continuation of Fig.~\ref{Realizations_on} (e).
S$\rightarrow$F$\rightarrow$S$\rightarrow$J transitions at bottleneck:
(a, c) Fragments of vehicle trajectories in different scales;
in (a) each vehicle trajectory is shown;
in (c) each 3rd vehicle trajectory is shown.
(b, d) Fragments of microscopic speeds; 
  bold dashed and dashed-dotted curves $v_{\rm max}(t)$ and  $v_{\rm min}(t)$
show, respectively,  the time-dependence of the maximum and minimum speeds on vehicle trajectories;
vehicle numbers in (b, d) are the same as those in     (a, c).
In (a, c), the beginning and   end of the on-ramp merging region are, respectively,
	$x_{\rm on}=$ 15 and $x^{\rm (e)}_{\rm on}=$ 15.3 km. 
  }
\label{On_S_F_J} 
\end{figure}

			  \begin{figure}
\begin{center}
\includegraphics*[width=8 cm]{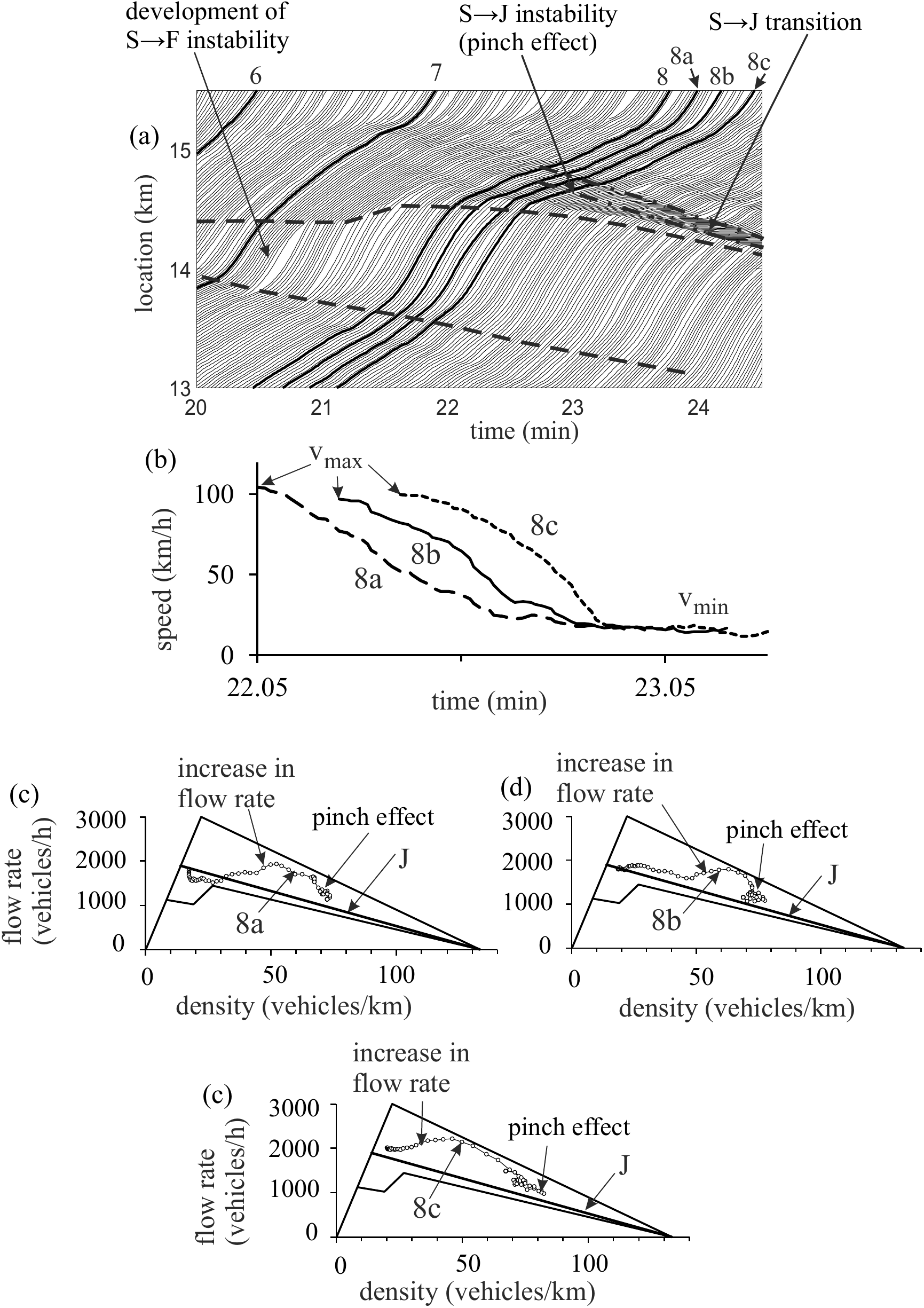}
\end{center}
\caption[]{Continuation of Fig.~\ref{On_S_F_J} (c):
(a) Fragment of vehicle trajectories;
vehicle numbers 6 and 7 are, respectively, the same as those   in  Fig.~\ref{On_S_F_J} (c).
(b) Fragments of microscopic speeds.
(c--e) Points along vehicle trajectories 8a (c), 8b (d), and 8c (e)  in the flow--density plane;
averaging over 10 vehicles upstream of the related vehicle for each time step (1 sec). 
Vehicle numbers in (b--e) are the same as those   in   (a).
In (a), the beginning and   end of the on-ramp merging region are, respectively,
	$x_{\rm on}=$ 15 and $x^{\rm (e)}_{\rm on}=$ 15.3 km. 
  }
\label{On_S_F_J_pinch_tr} 
\end{figure}

In a sequence of S$\rightarrow$F$\rightarrow$S$\rightarrow$J transitions at the bottleneck,
the first S$\rightarrow$F transition results from the development of 
an S$\rightarrow$F instability  (labeled by dashed curves in
  Fig.~\ref{On_S_F_J} (a, c))). A growing wave of the local speed increase caused by
	the S$\rightarrow$F instability is
		initiated by a speed peak that occurs at the bottleneck (labeled by 
$\lq\lq$speed peak" in Fig.~\ref{On_S_F_J} (b)). The physics of the speed peak
and of the development of the S$\rightarrow$F instability is qualitatively the same as that
found already in~\cite{Kerner2015B,Complex_Wave}. 

The   second S$\rightarrow$J transition in the
sequence of S$\rightarrow$F$\rightarrow$S$\rightarrow$J transitions 
  results from the occurrence of  the pinch effect in synchronized flow
(labeled by $\lq\lq$pinch effect" in 
Figs.~\ref{On_S_F_J_pinch_tr} (a, c--e)). The pinch effect is realized
 downstream of the wave of the local speed increase: During the vehicle deceleration
at the downstream front of this wave to a synchronized flow speed (see the speed reduction
of vehicles 8a, 8b, and 8c in Fig.~\ref{On_S_F_J_pinch_tr} (b)),
the flow rate increases (labeled by $\lq\lq$increase in flow rate" in 
Figs.~\ref{On_S_F_J_pinch_tr} (c--e)) and then the density in synchronized flow increases strongly
(labeled by $\lq\lq$pinch effect" in 
Figs.~\ref{On_S_F_J_pinch_tr} (c--e)).
The pinch effect causes the emergence of a growing narrow moving jam 
(S$\rightarrow$J instability)
(labeled by $\lq\lq$S$\rightarrow$J instability" in 
Fig.~\ref{On_S_F_J_pinch_tr} (a)). 
A growing narrow moving jam resulting from the pinch effect causes 
an S$\rightarrow$J transition (Figs.~\ref{On_S_F_J} (c)
and~\ref{On_S_F_J_pinch_tr} (a)).

The above physics of the  second S$\rightarrow$J transition in the
sequence of S$\rightarrow$F$\rightarrow$S$\rightarrow$J transitions at the bottleneck
is the same as that on the homogeneous road (Sec.~\ref{S-F-ini-S-J}). To explain this, we note that
after the growing wave of the local speed increase has occurred in synchronized flow at the bottleneck,
the subsequent development
of  the pinch effect with resulting the S$\rightarrow$J instability 
downstream of the wave is realized 
 far  upstream of the bottleneck. Therefore, this second S$\rightarrow$J instability
in the
sequence of S$\rightarrow$F$\rightarrow$S$\rightarrow$J transitions
does not  almost depend on the bottleneck existence.

\section{Discussion   \label{Dis_S}}

	\subsection{Effect of initial space gap on  phase transitions in synchronized flow on homogeneous road
	\label{Space_gap_S}}

	\subsubsection{Averaged speed  in synchronized flow
  \label{Speed_Sub}}
	
\begin{figure}
\begin{center}
\includegraphics*[width=7.5 cm]{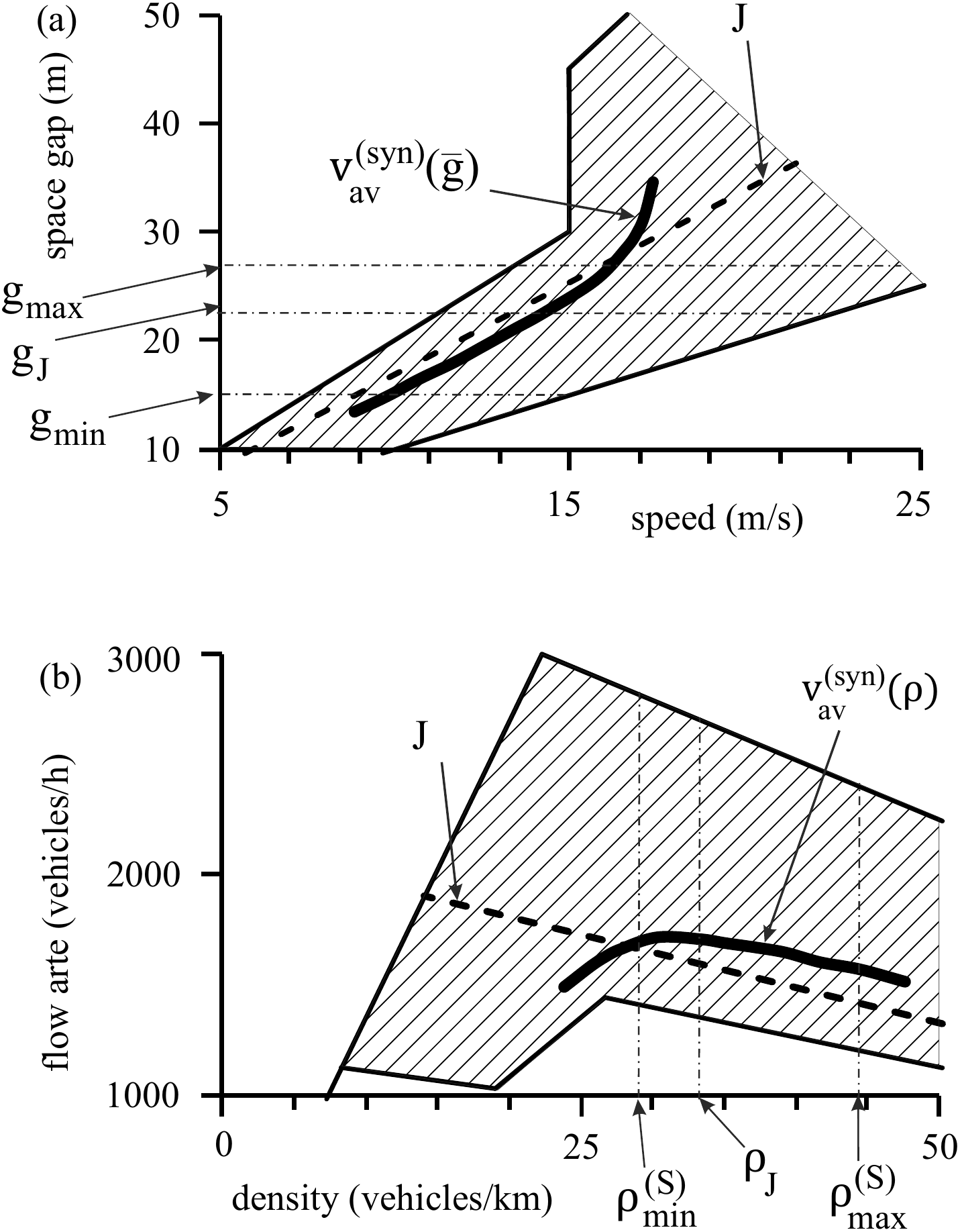}
\end{center}
\caption[]{A part of steady states of synchronized flow in the KKSW CA model (dashed regions) 
  in the space-gap--speed (a) and flow--density (b) planes together with
average speed in synchronized flow
(respectively, solid curves $v^{\rm (syn)}_{\rm av}(\overline g)$
and $v^{\rm (syn)}_{\rm av}(\rho)$) and
  with lines $J$ (dashed lines J).  $\rho^{\rm (S)}_{\rm max}=1000/(d+ g_{\rm min})$,
$\rho_{\rm J}=1000/(d+ g_{\rm J})$, $\rho^{\rm (S)}_{\rm min}=1000/(d+ g_{\rm max})$ (vehicles/km), 
$d$ is   vehicle length.
  }
\label{SpeedCurve} 
\end{figure}

	It has been found that when the initial   space-gap  $g_{\rm ini}$ between vehicles has been chosen
	(for example, $g_{\rm ini}=$ 19.5 m in Fig.~\ref{Realizations_Fig}), neither probabilities
	$P_{\rm S}$, $P_{\rm SF}$, $P_{\rm SJ}$ (Fig.~\ref{Prob_Fig}) nor qualitative features of phase transitions and resulting congested patterns (Fig.~\ref{Realizations_Fig}) depend on the value of 
	the initial synchronized flow speed $v^{\rm (syn)}_{\rm ini}$:
	Regardless of the choice of the initial synchronized flow speed, during a short time interval
	(about 1--2 min that can be considered negligible in comparison with the chosen time interval
	of traffic observation $T_{\rm ob}=$ 60 min)
	due to a random vehicle acceleration and deceleration
	the average synchronized flow speed tends to some almost time-independent
	value $v^{\rm (syn)}_{\rm av}$ for a given $g_{\rm ini}=\overline g$. This speed
	is a space-gap function $v^{\rm (syn)}_{\rm av}(\overline g)$
	(solid curve $v^{\rm (syn)}_{\rm av}(\overline g)$ in Fig.~\ref{SpeedCurve} (a)).

	Obviously that the function $v^{\rm (syn)}_{\rm av}(\overline g)$ exists
	within the space-gap range (\ref{Prob2_F})
	within which $P_{\rm S}>0$. However, it has been found that the function $v^{\rm (syn)}_{\rm av}(\overline g)$
	can be calculated even when $P_{\rm S}=0$ (Fig.~\ref{Prob_Fig} (a)), 
	however, only in some short space-gap ranges 
		related to $\overline g<  g_{\rm S}$
	and to $\overline g>  g_{\rm max}$
	that are outside   range (\ref{Prob2_F}).
	This is because condition $P_{\rm S}=0$ means that
  an S$\rightarrow$F transition   does occur in synchronized flow
	during the time interval $T_{\rm ob}$. As above-mentioned, this time interval is considerably longer than
	a short time interval of the reaching of the average synchronized flow speed
	$v^{\rm (syn)}_{\rm av}$. However, if either
	$\overline g$ becomes considerably smaller than $g_{\rm S}$
	or $\overline g$ becomes considerably larger than $g_{\rm max}$, the mean time delay of 
	the S$\rightarrow$F transition decreases to short enough values at which
	the average synchronized flow speed
	$v^{\rm (syn)}_{\rm av}$ cannot be calculated any more.
	This explains the existence of boundaries of  the
	space-gap function $v^{\rm (syn)}_{\rm av}(\overline g)$ in
	Fig.~\ref{SpeedCurve} (a) as well as of boundaries
  of  the
	density function $v^{\rm (syn)}_{\rm av}(\rho)$ in
	Fig.~\ref{SpeedCurve} (b).
		
	 \subsubsection{Phase transitions and resulting congested patterns
	\label{Space_gap_Sub}}

	As follows from Fig.~\ref{Prob_Fig}, for   
	space gap $g_{\rm ini}=$ 19.5 m   chosen
	in Fig.~\ref{Realizations_Fig}  
	probabilities $P_{\rm SF}=0.425$ and $P_{\rm SJ}=0.5$
	are close each other.
	The subsequent  strong spatiotemporal competition
	between S$\rightarrow$F and S$\rightarrow$J instabilities
	  results in the occurrence of the diverse variety of
		unknown features of congested traffic patterns 
		as well as effects like sequences
		of S$\rightarrow$F$\rightarrow$S$\rightarrow$J transitions
		and   S$\rightarrow$J$\rightarrow$S$\rightarrow$F transitions
		studied in Sec.~\ref{Probabilistic_S}.
		
		Here we discuss features of congested patterns
		when the spatiotemporal competition
	between S$\rightarrow$F and S$\rightarrow$J instabilities
	becomes weaker.
	This can be expected when due to the change in $g_{\rm ini}$ the difference between
	probabilities $P_{\rm SF}$ and $P_{\rm SJ}$ becomes larger.

	  \begin{figure}
\begin{center} 
\includegraphics*[width=8 cm]{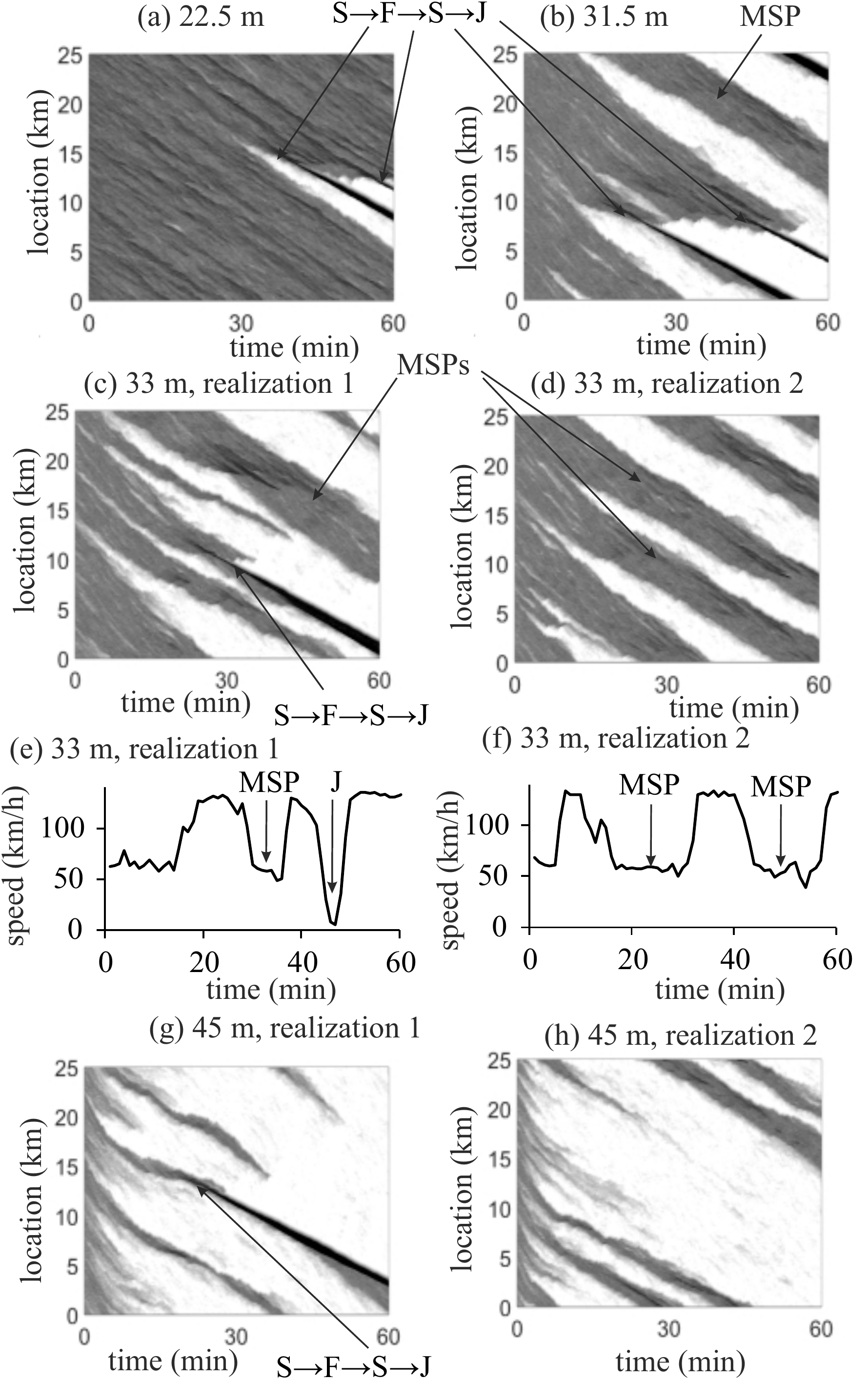}
\end{center}
\caption[]{Effect of the increase in initial
average space gap  between vehicles on   resulting congested patterns on homogeneous road:
(a) One of the realizations at $g_{\rm ini}=$ 22.5 m,
$v^{\rm (syn)}_{\rm ini}=$ 48.6 km/h.
(b) One of the realizations at $g_{\rm ini}=$ 31.5 m,
$v^{\rm (syn)}_{\rm ini}=$ 59.4 km/h.
(c, d) Two   realizations at $g_{\rm ini}=$ 33 m,
$v^{\rm (syn)}_{\rm ini}=$ 64.8 km/h.
(e, f) Time-dependencies of average speed (1 min averaged data) measured at virtual detectors at $x=$ 3 km for (c, d),
respectively.
(g, h) Two   realizations at $g_{\rm ini}=$ 45 m,
$v^{\rm (syn)}_{\rm ini}=$ 64.8 km/h. 
Other model parameters are the same as those  in Fig.~\ref{Realizations_Fig}.
 In (a--d, g, h), vehicle  speed data presented by  regions with 
variable shades of gray    (shades of gray vary from white to black when the speed 
decreases from 120 km/h (white) to 0 km/h (black)).
Values $(P_{\rm SF}, \ P_{\rm SJ})=$ (0.65, 0.08) (a), (1.0, 0) (b--h).
Arrows  S$\rightarrow$F$\rightarrow$S$\rightarrow$J 
 label some of the sequences of S$\rightarrow$F$\rightarrow$S$\rightarrow$J transitions.
MSP is a moving synchronized flow pattern.
  }
\label{Evolution_Fig} 
\end{figure}

\begin{figure}
\begin{center} 
\includegraphics*[width=8 cm]{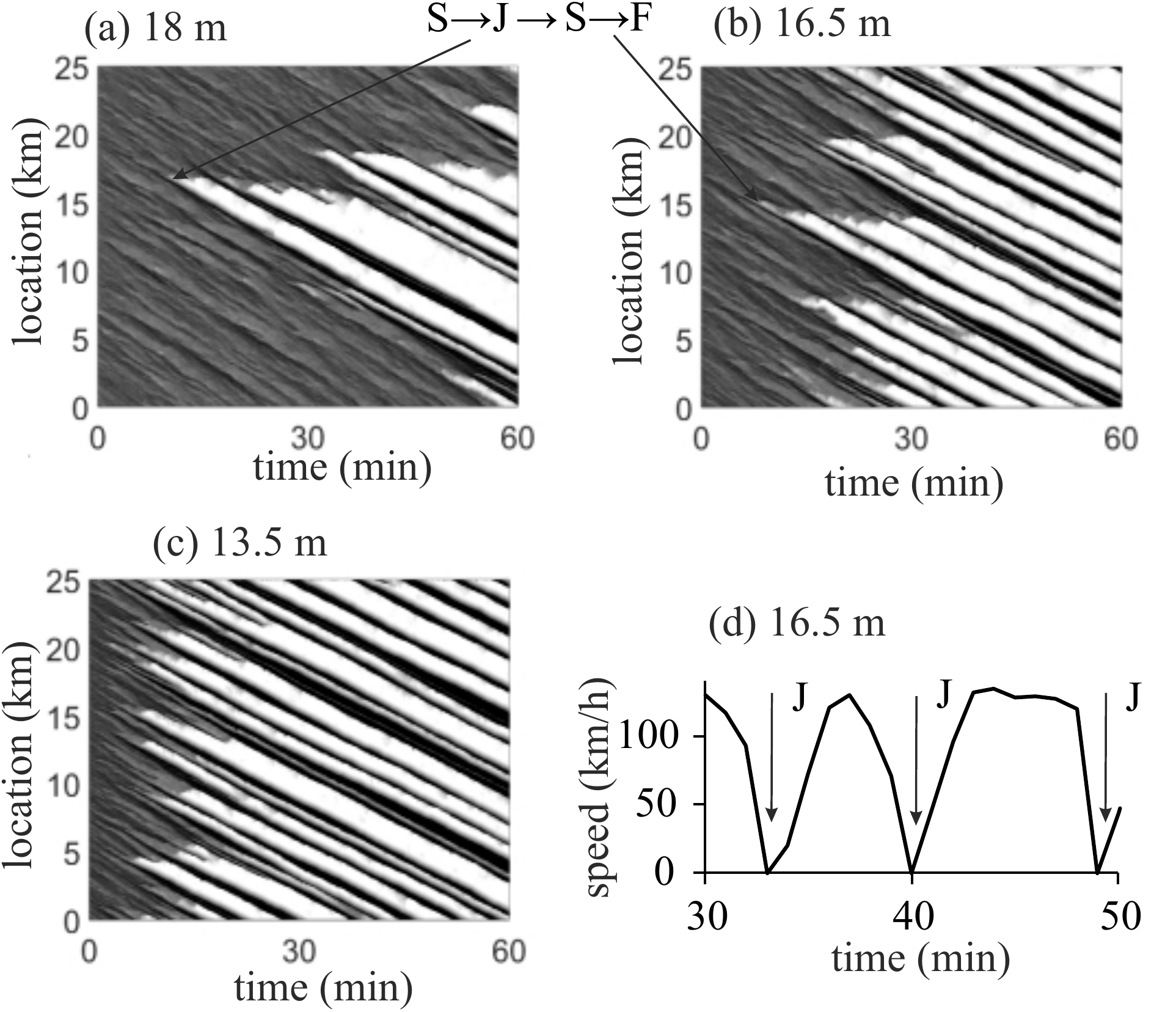}
\end{center}
\caption[]{Effect of the decrease in initial
average space gap  between vehicles on   resulting congested patterns on homogeneous road:
(a) One of the realizations at $g_{\rm ini}=$ 18 m,
$v^{\rm (syn)}_{\rm ini}=$ 43.2 km/h.
(b) One of the realizations at $g_{\rm ini}=$ 16.5 m,
$v^{\rm (syn)}_{\rm ini}=$ 37.8 km/h.
(c) One of the realizations at $g_{\rm ini}=$ 13.5 m,
$v^{\rm (syn)}_{\rm ini}=$ 32.4 km/h.
(d) Time-dependence of average speed (1 min averaged data) measured at a virtual detector at $x=$ 7 km for (b).  
Other model parameters are the same as those  in Fig.~\ref{Realizations_Fig}.
 In (a--c), vehicle  speed data presented by  regions with 
variable shades of gray    (shades of gray vary from white to black when the speed 
decreases from 120 km/h (white) to 0 km/h (black)).
Values $(P_{\rm SF}, \ P_{\rm SJ})=$ (0.2, 0.8) (a), (0.05, 0.95) (b), (0, 1.0) (c).
Arrows S$\rightarrow$J$\rightarrow$S$\rightarrow$F   label some of the sequences of S$\rightarrow$J$\rightarrow$S$\rightarrow$F transitions.
  }
\label{Evolution_Fig2} 
\end{figure}

	When $g_{\rm ini}$ increases, probability $P_{\rm SF}$ increases, whereas probability $P_{\rm SJ}$
	decreases (Fig.~\ref{Prob_Fig}).
Respectively, we have found that an S$\rightarrow$F   instability
governs mostly the emergence of the phases F and/or J in an initial synchronized flow
(Fig.~\ref{Evolution_Fig}). The larger the chosen average space gap $g_{\rm ini}$ is,
the more frequently one or a few moving synchronized flow patterns (MSP)  appear spontaneously
(some of the MSPs are labeled by $\lq\lq$MSP" in Fig.~\ref{Evolution_Fig}). MSPs emergence shown in
Fig.~\ref{Evolution_Fig} (b--d, g, h) is well-known effect~\cite{KernerBook}.
However, it has been unknown that due to the spatiotemporal competition
	between S$\rightarrow$F and S$\rightarrow$J instabilities
	the initial S$\rightarrow$F instability while causing a growing wave of a local speed increase in synchronized flow can result in an S$\rightarrow$J instability
	downstream of the wave. This effect (Sec.~\ref{S-F-ini-S-J})  
	results in S$\rightarrow$F$\rightarrow$S$\rightarrow$J transitions  even then, when probability
	$P_{\rm SJ}$ is very small (Fig.~\ref{Evolution_Fig} (a)) or $P_{\rm SJ}=0$
 (Fig.~\ref{Evolution_Fig} (b, c, g)).
	
	When $g_{\rm ini}$ decreases, probability $P_{\rm SF}$ decreases, whereas probability $P_{\rm SJ}$
	increases (Fig.~\ref{Prob_Fig}).
Respectively, we have found that an S$\rightarrow$J   instability
governs mostly the emergence of the phases J and/or F in an initial synchronized flow
(Fig.~\ref{Evolution_Fig2}). The smaller the chosen average space gap $g_{\rm ini}$ is,
the more frequently one or a few wide moving jams  appear spontaneously
(Fig.~\ref{Evolution_Fig2}). The wide moving jam emergence   is well-known effect~\cite{KernerBook}.
However, it has been unknown that due to the spatiotemporal competition
	between S$\rightarrow$F and S$\rightarrow$J instabilities
	the initial S$\rightarrow$J instability while causing a growing narrow moving jam
	  can result in an S$\rightarrow$F instability
	downstream of the moving jam. This effect (Sec.~\ref{S-J-S-F_trans})  
	results in S$\rightarrow$J$\rightarrow$S$\rightarrow$F transitions even then, when probability
	$P_{\rm SF}$ is very small (Fig.~\ref{Evolution_Fig2} (a, b)) or $P_{\rm SF}=0$
	 (Fig.~\ref{Evolution_Fig2} (c)).

	\subsection{Effect of on-ramp inflow rate
	on   phase transitions in synchronized flow at bottleneck  \label{Patterns_S}}
	
\begin{figure}
\begin{center} 
\includegraphics*[width=8 cm]{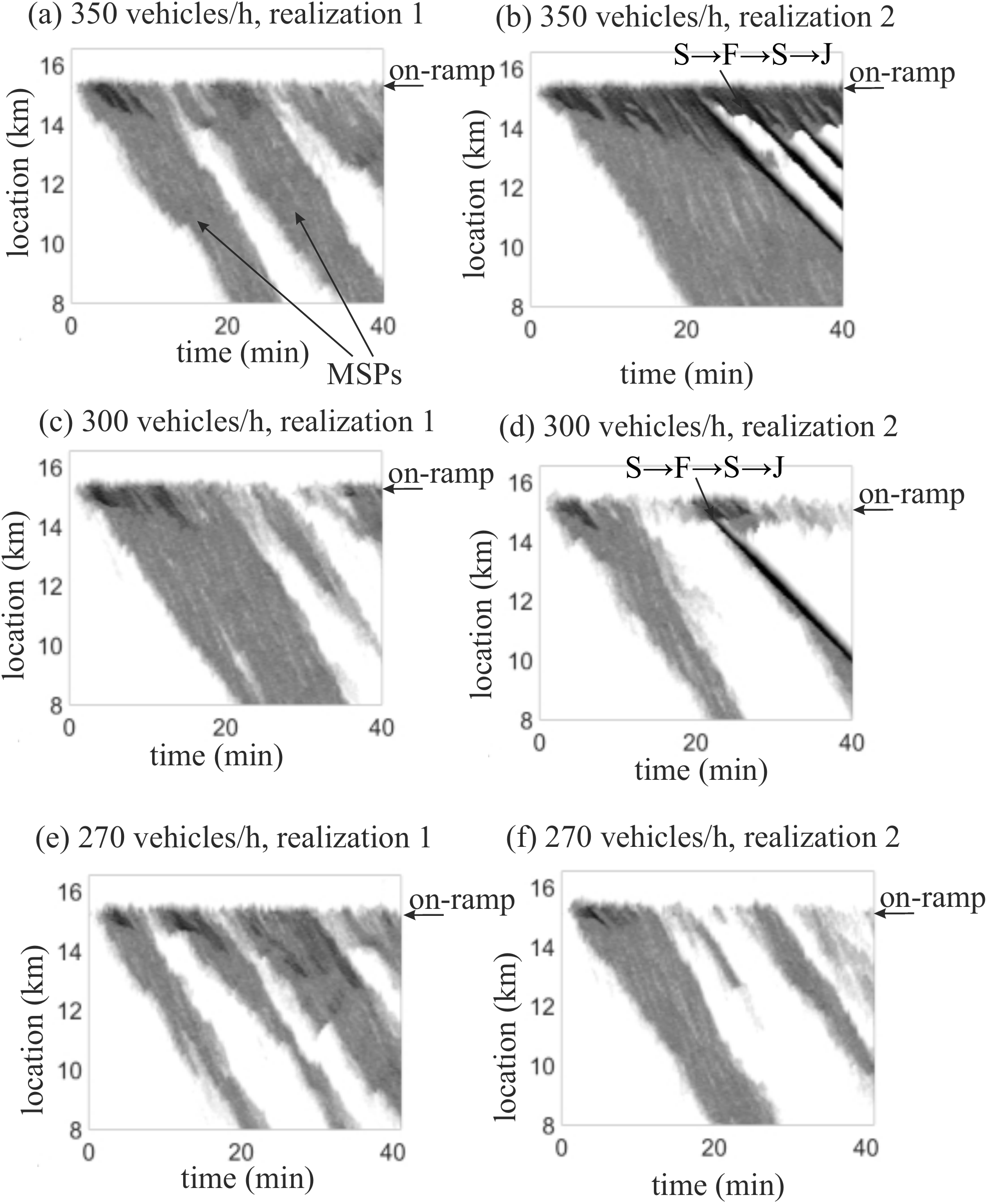}
\end{center}
\caption[]{Effect of the decrease in on-ramp inflow $q_{\rm on}$
  on   resulting congested patterns at bottleneck
	at the same other model parameters as those in Fig.~\ref{Realizations_on}:
(a, b) Two   realizations   at $q_{\rm on}=$ 350 vehicles/h. 
(c, d) Two   realizations   at $q_{\rm on}=$ 300 vehicles/h.
(e, f) Two   realizations   at $q_{\rm on}=$ 270 vehicles/h. 
 In (a--f), vehicle  speed data presented by  regions with 
variable shades of gray    (shades of gray vary from white to black when the speed 
decreases from 120 km/h (white) to 0 km/h (black)).
Values $(P_{\rm SF}, \ P_{\rm SJ})=$ (0.6, 0.125) (a, b), (1.0, 0) (c--f).
Arrows  S$\rightarrow$F$\rightarrow$S$\rightarrow$J 
 label some of the sequences of S$\rightarrow$F$\rightarrow$S$\rightarrow$J transitions.
  }
\label{Evolution_on} 
\end{figure}

\begin{figure}
\begin{center} 
\includegraphics*[width=8 cm]{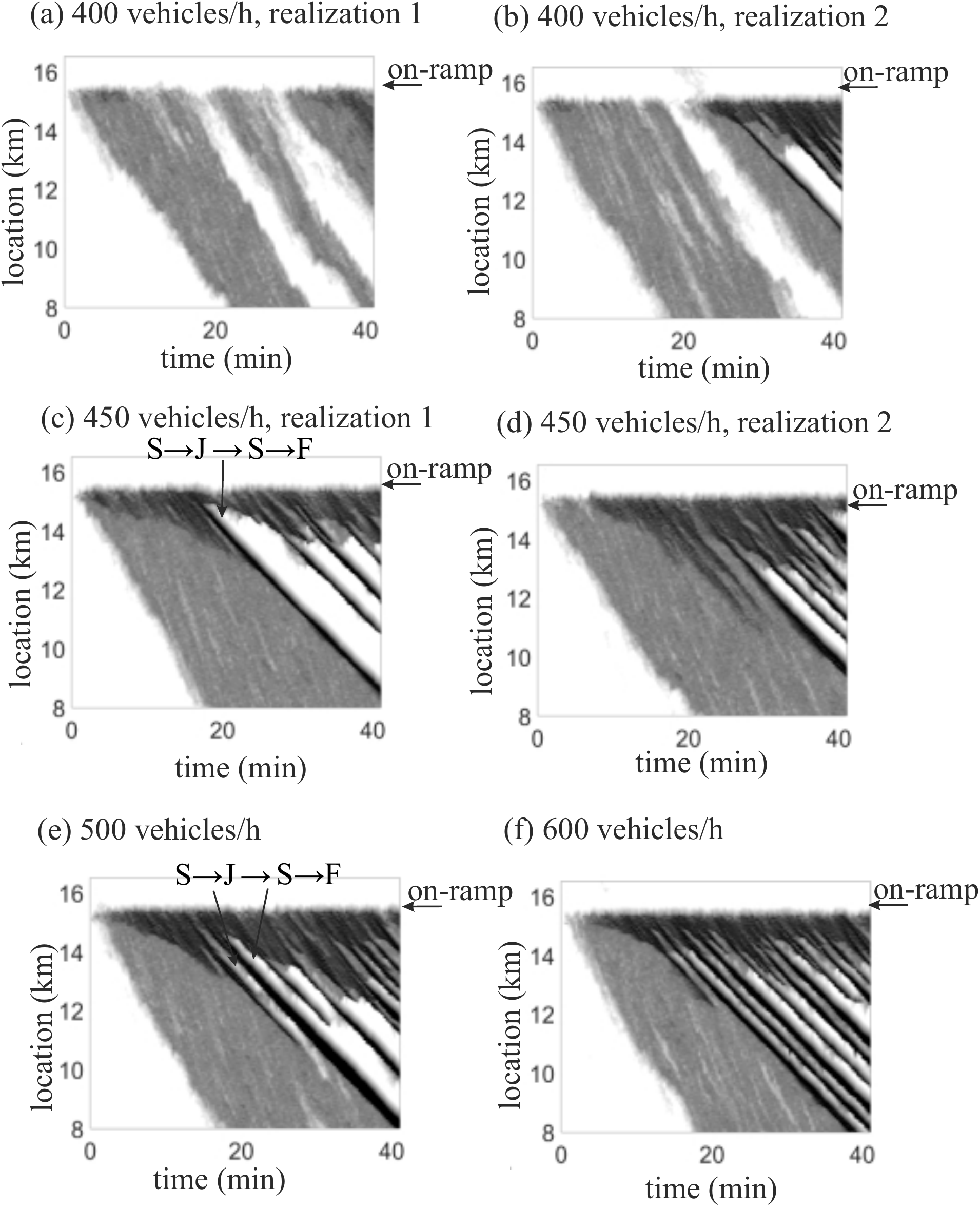}
\end{center}
\caption[]{Effect of the increase in on-ramp inflow $q_{\rm on}$
  on   resulting congested patterns at bottleneck
	at the same other model parameters as those in Fig.~\ref{Realizations_on}:
(a, b) Two   realizations   at $q_{\rm on}=$ 400 vehicles/h. 
(c, d) Two   realizations   at $q_{\rm on}=$ 450 vehicles/h.
(e) One of the   realizations   at $q_{\rm on}=$ 500 vehicles/h. 
(f) One of the   realizations   at $q_{\rm on}=$ 600 vehicles/h.
 In (a--f), vehicle  speed data presented by  regions with 
variable shades of gray    (shades of gray vary from white to black when the speed 
decreases from 120 km/h (white) to 0 km/h (black)).
Values $(P_{\rm SF}, \ P_{\rm SJ})=$ (0.3, 0.65) (a, b), (0.05, 0.95) (c, d), (0, 1.0) (e, f).
Arrows S$\rightarrow$J$\rightarrow$S$\rightarrow$F   label some of the sequences of S$\rightarrow$J$\rightarrow$S$\rightarrow$F transitions.
  }
\label{Evolution_on2} 
\end{figure}

	The effect of the average space gap between vehicles on 
	  phase transitions and resulting congested patterns on
the homogeneous road (Sec.~\ref{Space_gap_Sub}) is qualitatively similar
to the effect of the on-ramp inflow rate on 
	 phase transitions and resulting congested patterns at the bottleneck  
	(Figs.~\ref{Evolution_on} and~\ref{Evolution_on2}).
	
When $q_{\rm on}$ decreases, probability $P_{\rm SF}$ increases, whereas probability $P_{\rm SJ}$
	decreases (Fig.~\ref{Prob_on}).
Respectively, we have found that an S$\rightarrow$F   instability
governs mostly the emergence of the phases F and/or J in   synchronized flow at the bottleneck
(Fig.~\ref{Evolution_on}). As well-known~\cite{KernerBook}, the smaller   $q_{\rm on}$ is,  
the more frequently  MSPs  appear spontaneously
(some of the MSPs are labeled by $\lq\lq$MSP" in Fig.~\ref{Evolution_on}).  
However, it has been unknown that  due to the spatiotemporal competition
	between S$\rightarrow$F and S$\rightarrow$J instabilities
	the initial S$\rightarrow$F instability   can result in an S$\rightarrow$J instability
	upstream of the bottleneck. This effect (Sec.~\ref{On_S-F-J_on})     
	results in S$\rightarrow$F$\rightarrow$S$\rightarrow$J transitions even then, when probability
	$P_{\rm SJ}$ is very small (Fig.~\ref{Evolution_on} (b)) or $P_{\rm SJ}=0$
	 (Fig.~\ref{Evolution_on} (d)).
	
	When $q_{\rm on}$ increases, probability $P_{\rm SF}$ decreases, whereas probability $P_{\rm SJ}$
	increases (Fig.~\ref{Prob_on}).
Respectively, we have found that an S$\rightarrow$J   instability
governs mostly the emergence of the phases J and/or F in an initial synchronized flow at the bottleneck
(Fig.~\ref{Evolution_on2}). As well-known~\cite{KernerBook}, the larger $q_{\rm on}$ is,
the more frequently one or a few wide moving jams  appear spontaneously
(Fig.~\ref{Evolution_on2}).  
However, it has been unknown that  due to the spatiotemporal competition
	between S$\rightarrow$F and S$\rightarrow$J instabilities
	the initial S$\rightarrow$J instability  
	  can result in an S$\rightarrow$F instability
	upstream of the bottleneck. This effect (Sec.~\ref{On_S-J-S_on})  
	results in S$\rightarrow$J$\rightarrow$S$\rightarrow$F transitions even then, when probability
	$P_{\rm SF}$ is very small (Fig.~\ref{Evolution_on2} (c, d)) or $P_{\rm SF}=0$
   (Fig.~\ref{Evolution_on2} (e, f)).

	\subsection{F$\rightarrow$S$\rightarrow$F transitions before traffic breakdown and 
	phase transitions in synchronized flow at bottleneck  \label{FSF_S}}
 
	In the above study of statistical physics of synchronized flow at the bottleneck
	(Secs.~\ref{Probabilistic_On_S} and~\ref{Patterns_S}), we have
	  induced the synchronized flow through the use of the initial on-ramp inflow impulse.
		In real traffic, synchronized flow occurs often spontaneously
		due to a random time-delayed traffic breakdown (F$\rightarrow$S transition) at the bottleneck.
		In~\cite{Kerner2015B} it has been found that a random
		time delay $T^{\rm (B)}$  of this F$\rightarrow$S transition is governed by
		sequence(s) of F$\rightarrow$S$\rightarrow$F transitions that interrupt the  development
		of a congested pattern
		at the bottleneck:  Firstly, an F$\rightarrow$S transition
		has occurred at the bottleneck; then, in the emergent synchronized flow 
		 the S$\rightarrow$F instability is realized that leads to an S$\rightarrow$F transition; finally,  
		free flow returns at the bottleneck and the emergent synchronized flow dissolves
		(called as $\lq\lq$dissolving synchronized flow" at the bottleneck).
		Recently, sequences of F$\rightarrow$S$\rightarrow$F transitions before traffic breakdown
		predicted in~\cite{Kerner2015B} have indeed been observed in real field traffic data~\cite{Molzahn2017}. Thus, 
		a question arises: Do sequences of F$\rightarrow$S$\rightarrow$F transitions effect on statistical physical features of synchronized flow found above in this paper?

  \begin{figure}
\begin{center} 
\includegraphics*[width=8 cm]{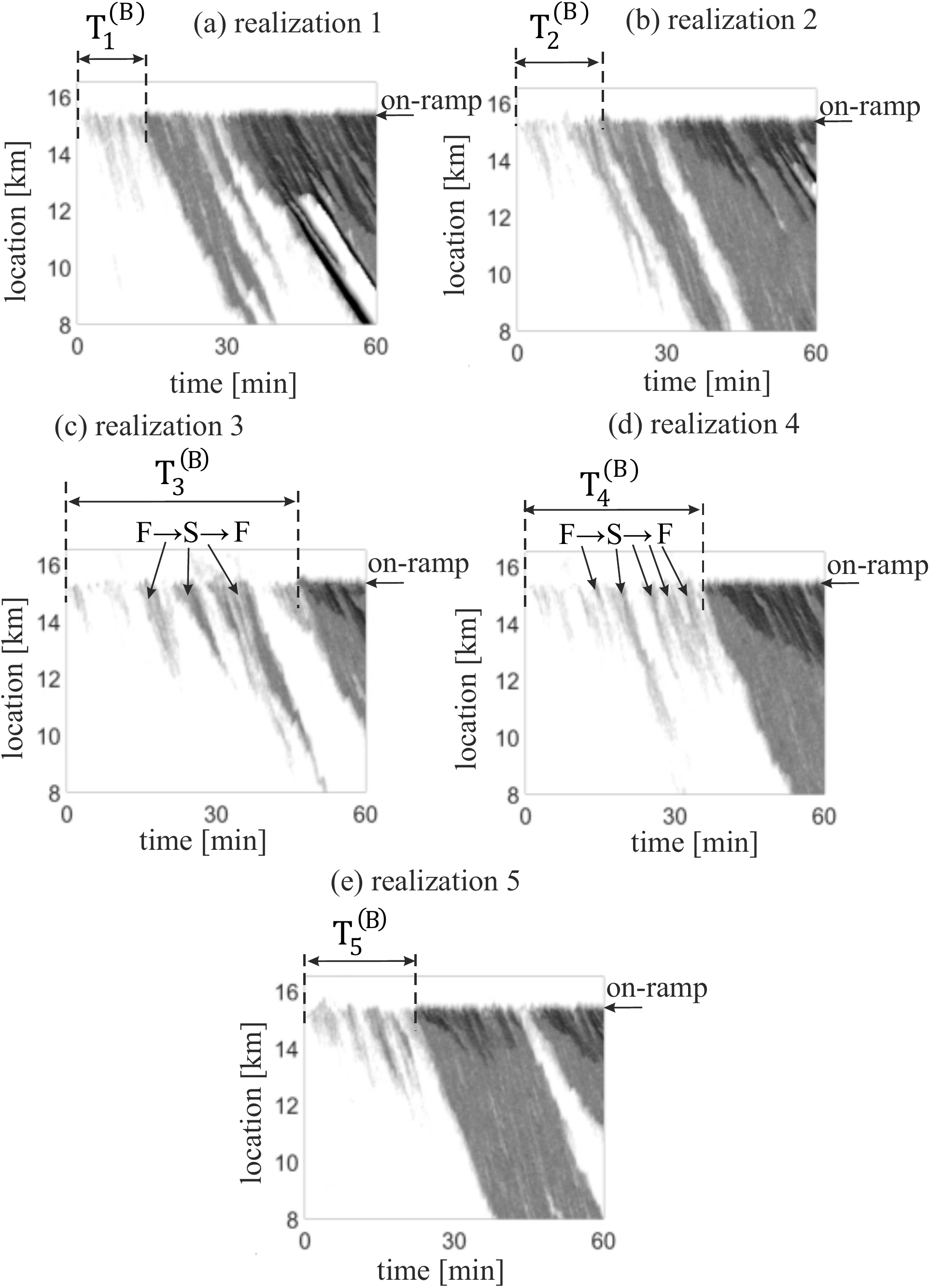}
\end{center}
\caption[]{F$\rightarrow$S$\rightarrow$F transitions and 
	phase transitions in synchronized flow at bottleneck: (a--e) five different realizations
	of spontaneous emergence of synchronized flow at the bottleneck. The flow rates
	$q_{\rm in}$, $q_{\rm on}$ and other
	  model parameters are the same as those in Fig.~\ref{Realizations_on}. The only one difference
	is that whereas in Fig.~\ref{Realizations_on} synchronized flow
	has been induced at the bottleneck, in (a--e) synchronized flow has emerged at the bottleneck spontaneously after
	a random time delay $T^{\rm (B)}_{i}$ ($i=$ 1, 2, ... 5 is the realization number)
	caused by F$\rightarrow$S$\rightarrow$F transitions before traffic breakdown.
 In (a--e), vehicle  speed data presented by  regions with 
variable shades of gray    (shades of gray vary from white to black when the speed 
decreases from 120 km/h (white) to 0 km/h (black)).
Arrows F$\rightarrow$S$\rightarrow$F   label some of the sequences of F$\rightarrow$S$\rightarrow$F transitions.
  }
\label{Realizations_FSF} 
\end{figure}

Simulations show that none of qualitative conclusions about statistical physics of 
synchronized flow at the bottleneck found above change, when
synchronized flow occurs spontaneously at the bottleneck. Some of simulation realizations
for this case made at the same model parameters as those in Fig.~\ref{Realizations_on} 
are presented in Fig.~\ref{Realizations_FSF}.
	
	\subsection{Conclusions  \label{Con_S}}

A spatiotemporal competition between    
S$\rightarrow$F and S$\rightarrow$J instabilities
is responsible for the following main statistical features of synchronized flow  revealed in the paper:

1.  There is a finite range of the initial space-gap between vehicles in synchronized flow
within which  during a chosen time for traffic observations
either synchronized flow persists with probability $P_{\rm S}$,
  or   firstly an S$\rightarrow$F transition  occurs
in synchronized flow with probability $P_{\rm SF}$,
or  else firstly an S$\rightarrow$J transition occurs
in synchronized flow with probability $P_{\rm SJ}$.

2. An initial S$\rightarrow$F   instability 
can cause the subsequent
S$\rightarrow$J instability downstream in synchronized flow while leading to the occurrence of
a sequence of S$\rightarrow$F$\rightarrow$S$\rightarrow$J transitions.

3. An initial S$\rightarrow$J   instability
can cause the subsequent
S$\rightarrow$F instability downstream in synchronized flow while leading to the occurrence of
a sequence of   S$\rightarrow$J$\rightarrow$S$\rightarrow$F transitions.
 
4. Each of the phase transitions in sequences of S$\rightarrow$F$\rightarrow$S$\rightarrow$J
and S$\rightarrow$J$\rightarrow$S$\rightarrow$F  transitions  
  exhibits the nucleation nature. This result
determines the spatiotemporal complexity of    traffic patterns.

5. At the same model parameters, there can be a large number of qualitatively different
 simulation realizations
in which different sequences of S$\rightarrow$F and S$\rightarrow$J instabilities
at random road locations are realized. The diverse variety of time-sequences 
of S$\rightarrow$F and S$\rightarrow$J instabilities occurring at random road locations
can cause  different    nucleation-interruption effects
  as well as different time-sequences of S$\rightarrow$F$\rightarrow$S$\rightarrow$J
and S$\rightarrow$J$\rightarrow$S$\rightarrow$F phase transitions.

6. Statistical features of vehicular traffic
found for a homogeneous road remain qualitatively for a road with a bottleneck.
In particular, 
  flow-rate dependencies of  probabilities
$P_{\rm S}$, $P_{\rm SF}$, and $P_{\rm SJ}$ 
 at the bottleneck are qualitatively the same
as  the space-gap dependencies of these probabilities found in the paper for the homogeneous road.

7. The main difference between the homogeneous road and the road with the bottleneck is that due to
a permanent non-homogeneity introduced by the bottleneck, nuclei for initial
S$\rightarrow$F and S$\rightarrow$J instabilities
appear mostly at the bottleneck rather than at random locations on the homogeneous road.

8. The phenomena of
the S$\rightarrow$F$\rightarrow$S$\rightarrow$J transitions and
the S$\rightarrow$J$\rightarrow$S$\rightarrow$F transitions caused by the competition of the S$\rightarrow$F and S$\rightarrow$J instabilities, which have been
found  in the paper both for the homogeneous road and for the road with the bottleneck
are common critical phenomena in synchronized flow. These
spatiotemporal traffic phenomena 
explain complex alternations of free flow regions with wide moving jams 
that occur when either the S$\rightarrow$J   instability
(Figs.~\ref{Realizations_Fig} (b, c) and~\ref{Realizations_on} (b, c)) or the   S$\rightarrow$F   instability
(Figs.~\ref{Realizations_Fig} (d, e) and~\ref{Realizations_on} (d, e)) is realized.

9. More complex spatiotemporal traffic phenomena have been found when in the same simulation realization
 a time-sequence of
S$\rightarrow$J and S$\rightarrow$F instabilities is realized. Examples
in which    firstly an S$\rightarrow$J instability occurs
and later an S$\rightarrow$F instability is realized are shown for 
  realization 6 for the homogeneous road
(Fig.~\ref{Realizations_Fig} (f)) as well as for realization 6 for the road with the bottleneck
(Fig.~\ref{Realizations_on} (f)). Other examples in which
  firstly an S$\rightarrow$F instability occurs
and later an S$\rightarrow$J instability is realized are shown for 
realization 7 for the homogeneous road
(Fig.~\ref{Realizations_Fig} (g)) as well as for realization 7 for the road with the bottleneck
(Fig.~\ref{Realizations_on} (g)).

\appendix
\section{KKSW CA model    \label{App_KKSW}}

 In all simulations we have used 
the KKSW CA three-phase traffic flow model~\cite{KKW,KKHS2013,KKS2014A}.  The physics of the KKSW CA model
has been considered in details in Appendix~B of the book~\cite{KernerBook3}.
In
 the KKSW CA model  for identical drivers and vehicles moving on   a single-lane road~\cite{KKS2014A}, the following  designations for main variables and vehicle parameters are used:
 $n=0, 1, 2, ...$ is the number of time steps; $\tau=1$ s is time step; $\delta x=1.5$ m is space step;
$x_{n}$ and $v_{n}$ are the coordinate and speed of the vehicle; time and space are measured in units of $\tau$ and $\delta x$, respectively;
$v_{\rm free}$ is the maximum speed in free flow; 
$g_{n}=x_{\ell,  n}-x_{n}-d$ is a space gap between two  vehicles following each other;
the lower index $\ell$ marks variables related to the preceding vehicle; $d$ is vehicle length; 
 $G_{n}$  is a synchronization space gap.

The KKSW CA model consists of the following sequence of rules~\cite{KKS2014A}:
\begin{description} 
\item {(a)} $\lq\lq$comparison of vehicle gap with the synchronization gap":  
 \begin{eqnarray} 
\label{conadaptation1}
\quad \mbox{if} \ g_{n} \leq G(v_{n}) \nonumber \\ 
 \quad \mbox{then  follow rules (b), (c) and skip rule (d),} 
\end{eqnarray}
 \begin{eqnarray}
\label{nonadaptation1}
 \quad \mbox{if} \ g_{n} > G(v_{n}) \nonumber \\
 \quad \mbox{then skip rules (b), (c) and follow rule (d),}
\end{eqnarray}
\item {(b)} $\lq\lq$speed adaptation within   synchronization   gap" is given by formula:
\begin{equation}
\label{adaptation_KKSW}
v_{n+1}=v_{n}+\mathrm{sgn}(v_{\ell,  n}-v_{n}),
\end{equation}
\item {(c)} $\lq\lq$over-acceleration through random acceleration within   synchronization   gap"
is given by formula
  \begin{eqnarray}
\label{Overacceleration1_KKW}
\mbox{if}   
  \ v_{n} \geq v_{\ell,  n}, \ \mbox{then
with probability} \ p_{\rm a}, \nonumber \\   
\quad  v_{n+1}=\min(v_{n+1}+1, \ v_{\rm free}),
\end{eqnarray}
\item {(d)} $\lq\lq$acceleration":
   \begin{equation}
\label{acceleration_KKSW}
v_{n+1}=\min(v_{n}+1, \ v_{\rm free}),
\end{equation}
\item {(e)} $\lq\lq$deceleration":
\begin{equation}
\label{Deceleration_KKSW}
v_{n+1}=\min (v_{n+1}, \ g_{n}),  
\end{equation}
\item {(f)} $\lq\lq$randomization"  is given by formula:
\begin{equation}
\label{Randomization_KKSW}
{\rm with} \ {\rm probability} \ p, \quad v_{n+1}=\max (v_{n+1}-1, \ 0),
\end{equation}
\item {(g)} $\lq\lq$motion" is described by formula:
\begin{equation}
\label{Motion_KKSW}
x_{n+1}=x_{n}+ v_{n+1}.
\end{equation}
\end{description}
Formula (\ref{Overacceleration1_KKW})  
 is applied,
when 
\begin{equation}
  r<p_{\rm a},
\label{rand_p1}
\end{equation}
formula (\ref{Randomization_KKSW})  
 is applied,
when 
\begin{equation}
p_{\rm a} \leq r<p_{\rm a}+p, 
\label{rand_p}
\end{equation}
where $p_{\rm a}+p\leq 1$; $r=rand()$ is a random value distributed uniformly between 0 and 1.
Probability of over-acceleration $p_{\rm a}$ in (\ref{Overacceleration1_KKW}) is chosen as 
the increasing speed function:
\begin{equation}
p_{\rm a}(v_{n})=p_{\rm a,1}+p_{\rm a,2}\max (0, \min (1, \ (v_{n}-v_{\rm syn})/\Delta v_{\rm syn})),
\label{p_acc}
\end{equation}
where $p_{\rm a,1}$, $p_{\rm a,2}$, $v_{\rm syn}$ and $\Delta v_{\rm syn}$ are constants.
 In  (\ref{conadaptation1}), (\ref{nonadaptation1}),
\begin{equation}
G(v_{n})=kv_{n}.
\label{S_Gap}
\end{equation}

The rules of vehicle motion  (\ref{nonadaptation1})--(\ref{S_Gap}) (without formula   (\ref{p_acc}))
have been formulated in the KKW (Kerner-Klenov-Wolf) CA model~\cite{KKW}.
In comparison with the KKW CA model~\cite{KKW}, we use in
 (\ref{Randomization_KKSW}), (\ref{rand_p})
  for probability $p$ formula  
 \begin{equation}
p=\left\{
\begin{array}{ll}
p_{2} &  \textrm{for  $v_{n+1}>v_{n}$}, \\
p_{3} &  \textrm{for  $v_{n+1} \leq v_{n}$},
\end{array} \right.
\label{p_MSP}
\end{equation}
which has been used in the KKSW CA model of Ref.~\cite{KKHS2013}.
The importance of formula (\ref{p_MSP}) is as follows.
This  rule of vehicle motion leads to a time delay in vehicle acceleration at the downstream front of synchronized flow.
In other words, this is an additional mechanism of time delay in vehicle acceleration in comparison with 
 a well-known  slow-to-start rule~\cite{Stoc2,Schadschneider_Book}:
\begin{equation}
p_{2}(v_{n})=\left\{
\begin{array}{ll}
p^{(2)}_{0} &  \textrm{for $v_{n}=0$},\\
p^{(2)}_{1} &  \textrm{for $v_{n}>0$} \\ 
\end{array} \right. 
\label{general}
\end{equation}
that is also used in the KKSW CA model. However, in the KKSW CA model
in formula  (\ref{general})  
probability  $p^{(2)}_{1}$ is   chosen to provide a delay in vehicle acceleration
only if the vehicle does not accelerate at previous time step $n$:
\begin{equation}
p^{(2)}_{1}=\left\{
\begin{array}{ll}
p^{(2)}_{2} &  \textrm{for $v_{n} \leq v_{n-1}$}, \\ 
0 &  \textrm{for $v_{n} > v_{n-1}$}. \\ 
\end{array} \right. 
\label{p_delay}
\end{equation} 
In (\ref{p_MSP})--(\ref{p_delay}), $p_{3}$, $p^{(2)}_{0}$, and $p^{(2)}_{2}$ are constants.
We also assume that
in (\ref{S_Gap})~\cite{KKW}
\begin{equation}
k(v_{n})=\left\{
\begin{array}{ll}
k_{1} &  \textrm{for $v_{n} > v_{\rm pinch}$},\\
k_{2} &  \textrm{for $v_{n} \leq v_{\rm pinch}$}, \\ 
\end{array} \right. 
\label{Pinch}
\end{equation}
where $v_{\rm pinch}$, $k_{1}$, and $k_{2}$ are constants ($k_{1}>k_{2}\geq 1$).

The rule of vehicle motion (\ref{p_MSP}) of the KKSW CA model~\cite{KKHS2013} together with formula (\ref{p_acc}) allows us to improve 
characteristics of synchronized flow patterns (SP) simulated with the KKSW CA model (\ref{nonadaptation1})--(\ref{Pinch}) for a single-lane road. Other physical 
features of the KKSW CA model have been explained in~\cite{KKHS2013}.

A model of an on-ramp bottleneck that has been used for simulations of single-lane road with the on-ramp 
bottleneck  as well as parameters of the model  are the same as those presented in
Appendix~B of the book~\cite{KernerBook3}.

 {\bf Acknowledgments:}
 We thank our partners for their support in the project $\lq\lq$MEC-View -- Object detection for automated driving based on Mobile Edge Computing",
    funded by the German Federal Ministry of Economic Affairs and Energy. 
I thank Sergey Klenov for discussions and help in simulations.


\begin{thebibliography}{8.}
\bibitem {Rand}
  	The only one difference in simulations of these realizations (Figs.~\ref{Realizations_Fig}
		and~\ref{Realizations_on})
	is a different choice    
	of the initial value (at $t=0$) of random function $rand()$
	in the KKSW CA model (see formulas (\ref{rand_p1}) and (\ref{rand_p}) in
		Appendix~\ref{App_KKSW}).
\bibitem{Prob_R}
  To find the probabilities of the   S$\rightarrow$F  
		and S$\rightarrow$J transitions in an initial   synchronized flow,
		only the most earlier occurring either the
			S$\rightarrow$F transition or the
			S$\rightarrow$J transition has been taken into account.
\end{thebibliography}
\end{document}